\def\Qem{{$Q_{\rm em}$}}

 \def\Z{{\bf Z}}

\def\EE{E$_8\times$E$_8^\prime$}

 \def\Z{{\bf Z}}

\def\EE{E$_8\times$E$_8^\prime$}
\def\Eo{E$_8$}

\def\Sp{${\cal S}$}
\def\Hvec{${\cal V}$}

\def\bftenh{${\bf 10}^H$}
\def\bftenbh{$\overline{\bf 10}^H$}
\def\bfst{{\bf 16}}

\documentclass{JHEP3}
\usepackage{epsf,epsfig,subfigure,graphicx,amsmath,amssymb}

\preprint{SNUTP 07-003 \\
KIAS-P07009}

\title{Superstring standard model from $\Z_{12-I}$ orbifold
compactification with and without exotics, and effective R-parity }
\author{Jihn E.  Kim and Ji-Hun Kim  \\
Department of Physics and Astronomy and Center for Theoretical
 Physics, Seoul National University, Seoul 151-747, Korea
\\E-mail: \email{jekim@phyp.snu.ac.kr,\ jihuni@phya.snu.ac.kr} }
\author{Bumseok Kyae \\
School of Physics, Korea Institute for Advanced Study, 207-43
Cheongryangri-dong, Dongdaemun-gu, Seoul 130-722, Korea
\\ E-mail: \email{bkyae@kias.re.kr} }

\abstract{We construct a supersymmetric standard model in the
context of the $\Z_{12-I}$ orbifold compactification of the
heterotic string theory. The gauge group is ${\rm SU(3)_c\times
SU(2)_L\times U(1)_Y\times U(1)^4\times[SO(10)\times U(1)^3]'}$.
We obtain three chiral families,
$3\times\{Q,d^c,u^c,L,e^c,\nu^c\}$, and Higgs doublets. There are
numerous neutral singlets many of which can have VEVs so that low
energy phenomenology on Yukawa couplings can be satisfied. In one
assignment (Model E) of the electroweak hypercharge, we obtain the
string scale value of $\sin^2\theta_W^0=\frac38$ and another
exactly massless {\it exphoton} (in addition to the photon)
coupling to exotic particles only. There are color triplet and
anti-triplet exotics, $\alpha$ and $\overline{\alpha}$, ${\rm
SU(2)_L}$ doublet exotics, $\delta$ and $\overline{\delta}$, and
${\rm SU(3)_c\times SU(2)_L}$ singlet but
$Y=\frac23,-\frac13,-\frac23,\frac13$ exotics,
$\xi,\eta,\overline{\xi}, \overline{\eta}$. We show that all these
vector-like exotics achieve heavy masses by appropriate VEVs of
neutral singlets. One can find an effective $R$-parity between
light (electroweak scale) particles so that proton and the LSP can
live sufficiently long. In another assignment (Model S) of the
electroweak hypercharge, there does not appear any exotic particle
but $\sin^2\theta_W^0=\frac{3}{14}$.}

\begin{document}

 \maketitle

\section{Introduction}

There has been numerous attempts to obtain  supersymmetric
standard models (SSM) from the orbifold compactification of
heterotic string \cite{DHVW:1985,INQ86,ChoiKimBk}. In the old
standardlike models, the attempts were just obtaining the standard
model gauge group and three families \cite{MSSMst}. In a recent
past, more ambitious attempts such as $\sin^2\theta_W=\frac38$
\cite{Kim03}, one pair of Higgs doublets \cite{Kim04}, and
neutrino masses \cite{Neutrino} were tried to be explained. More
recently, the Yukawa coupling structure has been looked for
\cite{Raby,Buch,KimKyaegut}. Among these, in particular we find
the GUT model of \cite{KimKyaegut} is satisfactory for the strong
CP solution via the QCD axion although a GUT scale axion decay
constant is needed \cite{flipax}, and for the approximate R-parity
violation \cite{KimRapp}.

From the proton longevity problem, the $R$-parity or matter parity
must be exact or feebly violated if it is an approximate one
\cite{ibross}. Otherwise, the string model construction must be
treated as an academic exercise. Even with a successful
$R$-parity, still there may be a good deal of phenomenological
problems to be overcome. Successful Yukawa coupling structure is
the next immediate concern in particle phenomenology. It is known
that the Yukawa coupling structure can be satisfied with the help
of numerous singlets \cite{Buch,KimKyaegut}. The next important
concern is the vacuum stabilization problem or the problem of flat
directions. But the vacuum stabilization problem is the most
difficult one to analyze. At present, we are not yet at the stage
to deal with this flat direction problem and we defer this flat
direction problem until we find a model satisfying other
phenomenological constraints. The approximate $R$-parity of
\cite{KimRapp} is the result of  GUT scale VEVs of \bftenh\ and
\bftenbh\ in the flipped SU(5) model. This hints that it may be
possible to obtain an exact $R$-parity if one succeeds in
obtaining an SSM   without such constraint on the GUT scale
VEVs.\footnote{If unlucky, such constraints will be replaced by
GUT scale constraints on singlet VEVs, which has to be checked
carefully.} Since the SSM through the flipped SU(5) was obtained
from a $\Z_{12-I}$ compactification, we look for a SSM directly in
the $\Z_{12-I}$ compactification. If found, the model is free from
the constraints of \bftenh\ and \bftenbh\ in the flipped SU(5)
model. But, then in a direct SSM construction one must check the
doublet-triplet splitting more carefully. A computer search of
SSMs is in principle possible but it is very difficult to put in
all the phenomenological requirements. At some stage a model by
model study is necessary. For example, we encounter a difficulty
of calculating the determinant of mass matrix of singlet exotics
in models with exotics whose number is much more than 10. The
determinant being zero up to some order of Yukawa couplings does
not necessarily mean that exotics do not obtain mass since still
higher orders might render a non-vanishing determinant.
Fortunately, for the $\Z_{12-I}$ compactification toward a direct
SSM, it has been possible to find out an SSM without the computer
search.

In this paper, we present an SSM in the $\Z_{12-I}$
compactification which can allow an {\it exact $R$-parity for low
energy} (electroweak scale) fields, which will be called an {\it
effective $R$-parity}. In the full theory, the $R$-parity is not
exact but the violation occurs through the type, $(heavy\
field)\to (light\ fields)$. With this kind of effective
$R$-parity, still the lightest supersymmetric particle (LSP) can
be a stable CDM candidate.

The  $R$-parity in the SO(10) GUT is achieved by different
assignments of quarks and leptons and Higgs doublets: in the  the
spinor \bfst\ for quarks and leptons and the vector {\bf 10} for
Higgs doublets. This kind of spinor-vector disparity can be
adopted in the {\it untwisted sector} of heterotic string also.
Let us consider only the \Eo\ part of the heterotic string
\cite{GroHar} for an illustration. The {\it untwisted sector}
massless matter spectrum in \Eo\ can be $P^2=2$ weights
distinguished by the spinor or the vector properties
\begin{align}
 {\cal S}: ([++++++++]) \quad
 {\cal V}: (\underline{\pm1~\pm1~0~0~0~0~0~0})\nonumber
\end{align}
where $\pm$ represents $\pm\frac12$, the notation [ ] means
including even number of sign flips inside the bracket, and the
underline means permutations of the entries on the underline. It
is obvious that cubic Yukawa couplings constructed with \Sp\ and
\Hvec\ respect a $\Z_2$ parity. But including matter from the
twisted sector, the study is more complex and we need the full
machinery of Yukawa couplings, including nonrenormalizable terms.
Here, the inclusion of neutral singlets, among which some needed
singlet VEVs can take the $\langle{\cal S}\rangle$ form, spoils
this idea of an exact $R$-parity. This needed singlet
$\langle{\cal S}\rangle$ is the reason that exact $R$-parity
models are extremely rare if not impossible. It is closely linked
to the assignment of the electroweak hypercharge $Y$. We will show
two interesting $Y$ assignments with the resulting physics such as
exotics, $\sin^2\theta_W$ and $R$-parity.

For the $R$-parity to be exact, it must be a subgroup of an
anomaly-free U(1) gauge group, i.e. it must be  a discrete gauge
symmetry \cite{discreteg}, otherwise large gravitational
corrections such as through wormhole processes may violate it.
Finding an anomaly free U(1) gauge symmetry direction whose $\Z_2$
subgroup is an $R$-parity is necessary for this purpose.

In Sec. \ref{sec:Z12Imodel}, we present an SSM from a $\Z_{12-I}$
compactification. Secs. \ref{sec:Vecexotics}--\ref{sec:Rparity}
discuss Model E. In Sec. \ref{sec:Vecexotics}, we list exotic
states which form vectorlike representations. We show how these
exotics obtain masses by VEVs of neutral singlets. In Sec.
\ref{sec:DFflat}, we discuss that there exist $D$- and $F$-flat
directions. In Sec. \ref{sec:Rparity}, we find a U(1) direction
whose $\Z_2$ subgroup can be used as an effective $R$-parity in
Model E. In Sec. \ref{sec:ModelS}, we discuss Model S. The
arguments on $D$- and $F$-flat directions and an effective
$R$-parity of Sec. \ref{sec:ModelS} are similar to those  given in
Sec. \ref{sec:DFflat} with minor corrections on the needed singlet
VEVs. Sec. \ref{sec:Conclusion} is a conclusion. In Appendix
\ref{App:A}, we list massless spectra according to the sectors. In
Appendix \ref{App:B}, we classify U(1) groups and find out the
anomalous U(1)$_A$ direction.

\section{SSM from $\Z_{12-I}$ compactification}\label{sec:Z12Imodel}

In \EE\ heterotic orbifold compactification, a model is completely
determined with (1) a twist vector $\phi$, which is associated
with the compactified 3 dimensional complex (or 6 dimensional
real) space, (2) a shift vector $V$ which is associated with the
16 dimensional ``gauge coordinate'' and (3) Wilson line introduced
in the compactified space. We employ the $\Z_{12-I}$ orbifold
specified with the twist vector
$\phi=(\frac{5}{12}~\frac{4}{12}~\frac{1}{12})$, and  take the
following shift vector $V$ and Wilson line $a_3$:
\begin{align}
&\phi= \textstyle (\frac{5}{12}~\frac{4}{12}~\frac{1}{12})\nonumber\\
 &V=\textstyle (\frac14~\frac14~\frac14~\frac14~\frac14~
;\frac{5}{12}~\frac{5}{12}~\frac{1}{12})
(\frac14~\frac34~0~;0^5)'\label{Z12Imodel} \\
&a_3=\textstyle
(\frac{2}{3}~\frac{2}{3}~\frac{2}{3}~\frac{-2}{3}~\frac{-2}{3}~;
\frac{2}{3}~0~\frac{2}{3}~)(0~\frac{2}{3}~\frac{2}{3}~;0^5)'\nonumber
.
\end{align}
They satisfy all the conditions required for modular
invariance~\cite{ChoiKimBk,Forste:2004ie}; $V_0^2-\phi^2=1$,
$a_3^2=4$, $V\cdot a_3=1$.  They give $V_+^2-\phi^2=7$ and
$V_-^2-\phi^2=3$, where $V_{0,+,-}=V+m_f a_3$ with $m_f=0,+1,-1$.

Low energy field spectrum in a model is determined with (1)
massless condition and (2) projection operator. The massless
conditions for left and right movers on an orbifold ${\bf Z}_N$
are
\begin{equation}
\begin{split}
\label{massless} &{\rm left\ movers}:\
\frac{(P+kV_f)^2}{2}+\sum_iN^L_i\tilde{\phi}_i -\tilde c_k=0 ,
\\
&{\rm right\ movers}:\
\frac{(s+k\phi)^2}{2}+\sum_iN^R_i\tilde{\phi}_i-c_k=0,
\end{split}
\end{equation}
where $k=0,1,2,\cdots,N-1$, $V_f=(V+m_fa_3)$, and $i$ runs over
$\{1,2,3,\bar{1},\bar{2},\bar{3}\}$.   Here $\tilde{\phi}_j\equiv
k\phi_j$ mod $Z$ such that $0<\tilde{\phi}_j\leq 1$, and
$\tilde{\phi}_{\bar{j}}\equiv -k\phi_j$ mod Z such that
$0<\tilde{\phi}_{\bar{j}}\leq 1$. If $k\phi_j$ is an integer,
$\tilde{\phi}_j=1$~\cite{Raby,Buch}. $N^L_i$ and $N^R_i$ indicate
oscillating numbers for left and right movers. It turns out that
$N^R_i=0$ generically for the massless right mover states in the
$\Z_{12-I}$ orbifold compactification. In Eqs. (\ref{massless}),
$P$ and $s$ [$\equiv (s_0,\tilde{s})$] are \EE\ and SO(8) weight
vectors, respectively. The values of $\tilde{c}_k$, $c_k$ are
found in Ref.~\cite{KimKyaegut}.

The multiplicity for a given massless state is calculated by the
projection operator \cite{ChoiKimBk,KimKyaegut},
\begin{eqnarray} \label{phase}
{\mathcal P}_k(f) = \frac{1}{NN_W}\sum_{l = 0 }^{N-1} \tilde{\chi}
( \theta^k , \theta^l ) e^{2 \pi i l\Theta_f}
 ,
\end{eqnarray}
where $f$ $(=\{f_0, f_+, f_-\})$ denotes twisted sectors
associated with $kV_f=kV$, $k(V+a_3)$, $k(V-a_3)$.  $N$ ($=12$ in
our case) is the order $N$ in the $\Z_N$ orbifold, and $N_W$ is
the order of the Wilson line, 3 in our case.  The phase $\Theta_f$
in Eq. (\ref{phase}) is given by
\begin{eqnarray}
\Theta_f = \sum_i (N^L_i - N^R_i ) \hat{\phi}_i - \frac{k}{2}
(V_f^2 - \phi^2 ) + (P+kV_f)\cdot V - (\tilde{s} + k \phi)\cdot
\phi,
\end{eqnarray}
where $\hat{\phi}_i = \phi_{i} ~{\rm sgn} (\tilde{\phi}_i)$. Here,
$\tilde{\chi}(\theta^k, \theta^l)$ is the degeneracy factor
summarized in Ref.~\cite{KimKyaegut}.  Note that ${\cal
P}_k(f_0)={\cal P}_k(f_+)={\cal P}_k(f_-)$ for $k=0,3,6,9$.

In addition, the left moving states in the $U$, $T_3$, $T_6$, and
$T_9$ sectors should satisfy \cite{Raby}
\begin{eqnarray}\label{condi3}
(P+kV)\cdot a_3=0~~{\rm mod}~~{\rm Z},~~~~{\rm for}\quad k=0, 3,
6, 9 .
\end{eqnarray}

\subsection{Massless spectra}

With the general formulae Eqs. (\ref{massless}), (\ref{phase}),
and (\ref{condi3}), and our choices Eq. (\ref{Z12Imodel}) the
massless spectra are calculated.
\subsubsection{Chirality and ${\cal N}=1$ SUSY}
The chirality and the number of supersymmetry (SUSY) ${\cal N}$ in
four dimensional spacetime (4D) after compactification are
determined by the massless right mover states.  Massless fermionic
states (``R-sector'') in the untwisted sector are represented by
the four component spinor $s=(s_0;\tilde s)=(\pm;\pm\pm\pm)$ with
even number of plus signs. Throughout this paper, $+$ ($-$)
denotes $\frac{+1}{2}$ ($\frac{-1}{2}$). $s_0$ determines the
chirality of a state. We define a state of $s_0=-$ ($+$) as the
left (right) handed state.  The corresponding bosonic states
(``NS-sector''), which also satisfy the massless condition for the
right mover, are obtained just by shifting the left-handed
[right-handed] fermionic state by $\tilde r_-=(-;-++)$ [$\tilde
r_+=(+;+--)$].

The ten dimensional SUSY generators are decomposed into
$Q_{(10)}=Q_{(4)} \otimes Q_{(6)}$. Under point group of the
orbifold, $Q_{(6)}$ transform as $Q_{(6)} \to \exp(2 \pi i s \cdot
\phi) Q_{(6)}$. The invariant component corresponds to the
unbroken supersymmetry generator in 4D. With $\phi=\textstyle
(\frac{5}{12}~\frac{4}{12}~\frac{1}{12})$, the solutions of $s$
satisfying $s \cdot \phi=$integer are only $(-;-++)$ and
$(+;+--)$, which give ${\cal N}=1$ SUSY because the number of
solutions counts the number of unbroken SUSY.

\subsubsection{Gauge symmetry and Weak mixing angle}
The gauge group and gauge quantum numbers are determined by the
massless left mover states. The root vectors of \EE\ satisfying
$P\cdot V=P\cdot a_3=0$~\cite{ChoiKimBk} are only
\begin{eqnarray}
(\underline{1,-1,0};0,0,0^3)(0^8)',\quad
(0,0,0;\underline{1,-1};0^8)(0^8)',\quad (0^8)(0^3;\underline{\pm
1,\pm1,0,0,0})' ,
\end{eqnarray}
where the underlined entries allow permutations. Thus the
resulting gauge group is
\begin{equation}
{\rm SU(3)_c\times SU(2)_L\times  U(1)_Y\times U(1)^4\times[SO(10)
\times U(1)^3]'} .
\end{equation}

Identification of the electroweak hypercharge is essential for the
assignment of SM fields, the GUT value of the weak mixing angle
$\sin^2\theta_W^0$, the appearance of exotics, and $R$ parity
assignments. In this paper, we present two identifications of the
electroweak hypercharge: (i) one with exotics and
$\sin^2\theta_W^0=\frac38$ and (ii) the other without exotics but
$\sin^2\theta_W^0=\frac{3}{14}$.  The electroweak hypercharge $Y$ is
defined as
\begin{itemize}
\item[{\bf }]
\begin{equation}
{\rm Model\ E:}\quad
Y=\textstyle(\frac13~\frac13~\frac13~\frac{-1}{2}~
\frac{-1}{2}~;0~0~0)(0~0~0;~0~0~0~0~0)' ,\label{ModelE}
\end{equation}
\item[{\bf }]
\begin{equation}
{\rm Model\ S:}\quad \tilde{Y}
=\textstyle(\frac13~\frac13~\frac13~\frac{-1}{2}~
\frac{-1}{2}~;0~0~0)(0~0~1;~0~0~0~0~0)' ,\label{ModelS}
\end{equation}
\end{itemize}
where Model E has exotics and $\sin^2\theta_W^0=\frac38$ and Model S
has only standard \Qem\ charges but $\sin^2\theta_W^0=\frac{3}{14}$.
Each assignment has its own merits and shortcomings. The hypercharge
$Y$ is orthogonal to every root vector of SU(3)$_c$, ${\rm
SU(2)_L}$, and SO(10)$'$. This operator turns out to give the
standard hypercharge assignments to the standard model (SM) chiral
fields viz. $Y(Q)=\frac16$, etc.

The current algebra in the heterotic string theory fixes the
normalization of $Y$. The $\sin^2\theta_W^0$ estimation is briefed
for Model E. Let us consider a properly normalized $Z$, which is
embedded in the string theory as
\begin{eqnarray}
Z = u\times Y=
u\times\left[\sqrt{\frac{2}{3}}~\frac{\vec{q}_3}{\sqrt{2}}
-\frac{\vec{q}_2}{\sqrt{2}}\right] ,
\end{eqnarray}
where $u$ indicates a normalization factor of $Y$, and $\vec{q}_3$
and $\vec{q}_2$ are orthonormal bases,
$\vec{q}_3=\frac{1}{\sqrt{3}}(1,1,1;0,0;0^3)(0^8)'$ and
$\vec{q}_2=\frac{1}{\sqrt{2}}(0,0,0;1,1;0^3)(0^8)'$. For $Z$ to be
embedded in the heterotic string theory, $u$ should be fixed such
that $u^2(\frac{2}{3}+1)=1$ or $u^2=\frac35$
\cite{ChoiKimBk,Ginsparg}. This hypercharge normalization leads to
a gauge coupling normalization $g_1^2=\frac53g_Y^2$, where $g_1$
is unified at the string scale with the non-Abelian gauge
couplings such as ${\rm SU(2)_L}$ gauge coupling $g_2$. Thus, in
Model E the weak mixing angle at the string scale is
\begin{eqnarray}
 \sin^2\theta_W^0=\frac{1}{1+({g_2^2}/{g_Y^2})}=\textstyle\frac38 .
\end{eqnarray}
The same kind of calculation gives $\sin^2\theta_W^0=\frac{3}{14}$
in Model S.

Since $\tilde Y$ in Model S is obtained by adding a U(1)$_6$
generator belonging to E$_8'$, in the bulk of the paper (except
Sec. \ref{sec:ModelS}) we present quantum numbers of Model E and
an effective $R$-parity. Then, in Sec. \ref{sec:ModelS} we present
Model S.

\subsubsection{Chiral matter}
The matter spectra appear from the untwisted and twisted sectors.
All matter fields in this model are tabulated in Tables
\ref{tb:untwistedvi}--\ref{tb:T5} in Appendix \ref{App:A}.
Depending on the values of $P\cdot V$, the origins of the fields
are denoted by $U_1, U_2, U_3$ for the untwisted sector fields. We
name the twisted sector associated with $kV_f=(V+m_fa_3)$
``$T_k^{m_f}$'' with superscripts $0,+,-$ (except for
$T_3,T_6,T_9$).  For modular invariance, all these sectors should
be considered.

In a ${\bf Z}_N$ orbifold compactification, the anti-particle
states (${\cal CTP}$ conjugations) of particle states in a
$T_k^{m_f}$ sector are, in general, found from the $T_{N-k}^{m_f}$
sector. In the ${\bf Z}_{12-I}$ case, the untwisted sector $U$ and
 $T_3$, $T_6$, $T_9$ sectors provide both left and right
chirality states. In particular, the $U$ and $T_6$ sectors contain
particle states and their corresponding anti-particles states.  On
the other hand,  $T_1$, $T_2$, $T_4$, $T_7$ ($T_{11}$, $T_{10}$,
$T_8$, $T_5$) sectors allow only left (right) chirality states.

As seen in the Tables \ref{tb:untwistedvi}--\ref{tb:T5}, this
model allows three families of SSM matter fields from the
$U_{1,3}$ and $T_4^0$ sectors.  The other fields including the
electroweak Higgs are vectorlike under the SM gauge symmetry:
\begin{eqnarray}
3\times \{Q,~d^c,~u^c,~L,~e^c,~\nu^c\}~~ +~~ {\rm
vectorlike~fields~(including~MSSM~Higgs)}.
\end{eqnarray}
The key representations of this SSM are
\begin{align}
{\rm matter}:&\left\{\begin{array}{l} Q=({\bf 3}, {\bf
2})_{\frac{1}{6}},\  d^c=({\bf 3}^*,
{\bf 1})_{\frac{1}{3}},\  u^c=({\bf 3}^*, {\bf 1})_{\frac{-2}{3}},\\
 L= ({\bf 1}, {\bf 2})_{\frac{-1}{2}},\ e^c=
({\bf 1}, {\bf 1})_{1},\ \nu^c= ({\bf 1}, {\bf 1})_{0},\
\end{array}\right.
\label{eq:matter}\\
{\rm Higgs}:~&\left\{\begin{array}{l} H_u= ({\bf 1}, {\bf
2})_{\frac{1}{2}},\ H_d= ({\bf 1}, {\bf 2})_{\frac{-1}{2}},\
\quad{\rm electroweak\ scale}\\
{\bf 1}_0, \quad{\rm string\ scale} .
 \end{array}\right.\label{eq:Higgs}
\end{align}
In this model, there are vectorlike $D$ and $\overline{D}$ (color
triplet and antitriplet fields) which carry the familiar $d$-type
quark charge \Qem$=\mp\frac13$, respectively.

We observe also that there are states with exotic electromagnetic
charges (exotics) from the $T_{k}^{\pm}$ ($k=1,2,4,7$) sectors.
All {\it color exotics} are SU(3)$_c$ triplets and antitriplets
and carry \Qem $=0,\pm\frac13$.  The SU(2) doublet exotics or
simply {\it doublet exotics} carry $Y=\pm\frac16$ whose components
carry again \Qem $=\pm\frac23,\pm\frac13$. The ${\rm SU(3)_c\times
SU(2)_L}$ {\it singlet exotics} carry \Qem
$=\pm\frac23,\pm\frac13$. {\it All these exotics form vectorlike
representations under the SM gauge symmetry.}
\footnote{Since all the SSM matter fields arise from the $U$ and
$T_4^0$ sectors, while all the exotics are only from the twist
sectors associated with Wilson line $T_{k}^{\pm}$ ($k=1,2,4,7$), 3
families of SSM matter fields are relatively easily obtained even
with other choices of Wilson line. Indeed, a large class of models
with $\frac14$ as the first five entries in the shift vector $V$
and with a proper Wilson line can give ${\rm
sin^2}\theta_W=\frac38$ and 3 families of the SSM matter fields.
However, it is non-trivial to construct a model such that all
exotics form vectorlike representations under the SM gauge
symmetry.}
The mass scales of these vectorlike representations are near the
string scale if the needed neutral singlets develop string scale
VEVs. We will comment more on this later.

\begin{table}[t]
\begin{center}
\begin{tabular}{|c|c|c|c|}
\hline  Visible states & SM notation& $\Gamma$ & $\Gamma'$
\\
\hline
 $(\underline{++-};\underline{+-};+++)(0^8)'$ & $Q(U_1)$ &--1 & +1
\\
$(\underline{+--};--;+++)(0^8)'$ & $d^c(U_3)$ & --1 & +1
\\
$(\underline{+--};++;+--)(0^8)'$ & $u^c(U_3)$ &--1 & $-3$
\\
$(---;\underline{+-};+--)(0^8)'$ & $L(U_1)$ & --1 & $-3$
\\
$(+++;--;-+-)(0^8)'$ & $e^c(U_3)$ & +5 & +5
\\
$({+++;++};+++)(0^8)'$ & $\nu^c(U_3)$ &--1 & +1
\\
$(0~0~0;\underline{-1~0};-1~0~0)(0^8)'$ & $H_u(U_2)$ & +2 & +2
\\
$(0~0~0;\underline{1~0};0~0~1)(0^8)'$ & $H_d(U_2)$ & --4 & $-2$
\\
 $(\underline{++-};\underline{+-};\frac16~\frac16~
 \frac{-1}{6})(0^8)'$ & $2\cdot Q(T_4^0)$ &+1 & +1
 \\
$(\underline{+--};{--};\frac16~\frac16~
 \frac{-1}{6})(0^8)'$ & $2\cdot d^c(T_4^0)$ &+1 & +1
\\
 $(\underline{+--};{++};\frac16~\frac16~
 \frac{-1}{6})(0^8)'$ & $2\cdot u^c(T_4^0)$ & $-3$ & $-3$
\\
$({---};\underline{+-};\frac16~\frac16~
 \frac{-1}{6})(0^8)'$ & $2\cdot L(T_4^0)$ & $-3$ & $-3$
\\
$({+++};{--};\frac16~\frac16~
 \frac{-1}{6})(0^8)'$ & $2\cdot e^c(T_4^0)$ &+5 & +5
\\
$({+++};{++};\frac16~\frac16~
 \frac{-1}{6})(0^8)'$ & $2\cdot \nu^c(T_4^0)$ &+1 & +1
\\
$(\underline{1,0,0};0~0;\frac{-1}{3}~\frac{-1}{3}~\frac{1}{3})(0^8)'$
& $3\cdot \overline{D}_{1/3}(T_4^0)$ & $\fbox{+2}$ & $\fbox{+2}$\\
[0.1em]
$(\underline{-1,0,0};0~0;\frac{-1}{3}~\frac{-1}{3}~\frac{1}{3})(0^8)'$
& $2\cdot {D}_{-1/3}(T_4^0)$ &$\fbox{$-2$}$ & $\fbox{$-2$}$\\
$({0,0,0};\underline{-1~0};\frac{-1}{3}~
\frac{-1}{3}~\frac{1}{3})(0^8)'$ & $2\cdot H_u(T_4^0)$ &+2 & +2\\
$({0,0,0};\underline{1~0};\frac{-1}{3}~\frac{-1}{3}~\frac{1}{3})(0^8)'$
& $3\cdot H_d(T_4^0)$ &--2 & $-2$\\
$(\underline{1,0,0};{0~0};0^3)(\frac{-1}{2}~\frac12~0;0^5)'$ &
$3\cdot \overline{D}_{1/3}(T_6)$ & $\fbox{+2}$ & $\fbox{$+2$}$
\\ [0.1em]
$(\underline{-1,0,0};{0~0};0^3)(\frac{1}{2}~\frac{-1}{2}~0;0^5)'$
& $3\cdot {D}_{-1/3}(T_6)$ & $\fbox{$-2$}$ & $\fbox{$-2$}$\\
 $({0,0,0};\underline{-1~0};0^3)(\frac{-1}{2}~\frac12~0;0^5)'$ &
$2\cdot H_u(T_6)$ &+2 & +2\\
 $({0,0,0};\underline{1~0};0^3)(\frac{1}{2}~\frac{-1}{2}~0;0^5)'$ &
$2\cdot H_d(T_6)$ &$-2$ & $-2$\\

 $(\underline{\frac34\frac{-1}{4}\frac{-1}{4}};{\frac{-1}{4}
  \frac{-1}{4}};\frac{1}{4}\frac{1}{4}\frac{1}{4})
  (\frac{3}{4}\frac{1}{4}~0;0^5)'$ &
$\overline{D}_{1/3}(T_3)$ &${1}$ & \fbox{+2}\\
  $(\underline{\frac{-3}{4}\frac{1}{4}\frac{1}{4}};{\frac{1}{4}
  \frac{1}{4}};\frac{-1}{4}\frac{-1}{4}\frac{-1}{4})
  (\frac{-3}{4}\frac{-1}{4}~0;0^5)'$ &
$2\cdot {D}_{-1/3}(T_9)$ &${-1}$ & \fbox{$-2$}\\
  $({\frac{1}{4}\frac{1}{4}\frac{1}{4}};\underline{\frac{-3}{4}
  \frac{1}{4}};\frac{-1}{4}\frac{-1}{4}\frac{-1}{4})
  (\frac{1}{4}\frac{3}{4}~0;0^5)'$ &
$2\cdot H_u(T_9)$ &${+4}$ & +3\\
  $({\frac{-1}{4}\frac{-1}{4}\frac{-1}{4}};\underline{\frac{3}{4}
  \frac{-1}{4}};\frac{1}{4}\frac{1}{4}\frac{1}{4})
  (\frac{-1}{4}\frac{-3}{4}~0;0^5)'$ &
$H_d(T_3)$ &${-4}$ & $-3$
\\
\hline
\end{tabular}
\end{center}
\caption{Standard charge left-handed (L) chiral fields. The
multiplicity is shown as the coefficients of representations. +
and -- represent $+\frac12$ and $-\frac12$, respectively.  Neutral
singlets  are listed in the following table. $\overline{D}_{1/3}$
and ${D}_{-1/3}$ in $T_4^0$ and $T_6$ have unconventional
$\Gamma$s, not mixing with $d$ and $d^c$ with an exact
parity.}\label{tb:standardch}
\end{table}

In Table \ref{tb:standardch}, we list particles carrying familiar
\Qem\ charges.  In addition, we list neutral singlets in Table
\ref{tb:Singneutral}. Some of these neutral singlets are required
to have string scale VEVs in order to break extra U(1)s and give
masses to the exotics.

In the $T_3$ and $T_9$ sectors as shown in Table \ref{T3states} of
Appendix, there are three ${\bf 10}'$s of ${\rm SO(10)}'$.  In
this model, the hidden sector confining group is ${\rm SO(10)'}$.
We assume that some of three ${\bf 10}'$s of ${\rm SO(10)'}$
obtain VEVs and break ${\rm SO(10)'}$ to a smaller nonabelian
group so that its confining scale is at the intermediate scale.
The gaugino condensation at this intermediate scale would break
the ${\cal N}=1$ SUSY.

\subsection{Yukawa couplings} \label{rules}

To study Yukawa couplings in orbifold compactification, we need to
know the $H$-momentum of a state in a sector. Neglecting the
oscillator numbers, the $H$-momenta of states, $H_{\rm mom,0}$
[$\equiv(\tilde s+k\phi+\tilde r_-)$] are
\begin{align}
&U_1: (-1,0,0),\quad U_2: (0,1,0),\quad U_3:
(0,0,1),\nonumber\\
&\textstyle T_1:(\frac{-7}{12},\frac{4}{12},\frac1{12}),\quad
 T_2:(\frac{-1}{6},\frac46,\frac16),\quad T_3:
 (\frac{-3}{4},0,\frac{1}{4}),\nonumber\\
&\textstyle
 T_4:(\frac{-1}{3},\frac13,\frac13),\quad
\left\{T_5:(\frac{1}{12},\frac{-4}{12},\frac{-7}{12})\right\},
\quad
T_6:(\frac{-1}{2},0,\frac12),\\
&\textstyle T_7:(\frac{-1}{12},\frac{4}{12},\frac{7}{12}),\quad
T_9:(\frac{-1}{4},0,\frac{3}{4}) , \nonumber
\end{align}
from which $T_5$ will not be used since the chiral fields there
are right-handed while the other fields are represented as
left-handed. With oscillators, the $H$-momentum [$\equiv
(R_1,R_2,R_3)$] are
\begin{equation}
(H_{\rm mom})_j= (H_{\rm mom,0})_j  - (N^L)_j + (N^L)_{\bar{j}} ~
,\quad j=1,2,3 . \label{Hmom}
\end{equation}

The superpotential terms  are obtained by examining vertex
operators satisfying the orbifold conditions \cite{ChoiKimBk}. It
can be summarized as the following selection rules:
\begin{itemize}
\item[(a)] Gauge invariance. \item[(b)] $H$-momentum conservation
with $\phi=
  \left(\frac{5}{12}, \frac{4}{12}, \frac{1}{12} \right)$,
\begin{eqnarray}
\sum_z R_1 (z) = -1 {\rm~ mod~} 12 , \quad \sum_z R_2 (z) = 1
{\rm~ mod~} 3,\quad \sum_z R_3 (z) = 1 {\rm~ mod~}
12,\label{Hconsv}
\end{eqnarray}
where $z(\equiv A,B,C,\dots)$ denotes the index of states
participating in a vertex operator. \item[(c)] Space group
selection rules:
\begin{eqnarray}
&& \sum_z k(z) = 0 {\rm~ mod~} 12,\label{modinvk} \\
&& \sum_z \left[ km_f \right] (z) = 0 {\rm~ mod~}
3.\label{modinva}
\end{eqnarray}
\end{itemize}

If some singlets obtain string scale VEVs, however, the condition
(b) can be merged into Eq. (\ref{modinvk}) in (c). Our strategy to
see this is to construct composite singlets (CS) which have
$H$-momenta, (1,0,0), $(-1,0,0)$, (0,1,0), (0,$-1$,0), (0,0,1),
(0,0,$-1$), using only singlets developing VEVs of order $M_{\rm
string}$. Then, with any integer set $(l,m,n)$, we can attach an
appropriate number of CSs to make the total $H$-momentum be
$(-1,1,1)$. Indeed, it is possible to construct such CSs, with the
singlets defined in Table \ref{tb:Singneutral}:
\begin{align}\label{CSHmom}
&[S_1S_8^{(1)}S_{10}] [S_4^{(3)}S_7^{(1)}S_{12}]
[S_4^{(1)}S_7^{(3)}S_{12}]:\ (1,0,0),\nonumber \\
&[S_1S_8^{(1)}S_{10}] [S_4^{(3)}S_7^{(1)}S_{12}]
[S_1S_8^{(3)}S_{10}]:\ (-1,0,0), \nonumber\\
&[S_1S_8^{(1)}S_{10}] [S_4^{(3)}S_7^{(1)}S_{12}]
[S_1S_8^{(3)}S_{10}] [S_4^{(1)}S_7^{(3)}S_{12}]:\ (0,1,0) ,\\
&[S_1S_8^{(1)}S_{10}][S_4^{(3)}S_7^{(1)}S_{12}]:
\ (0,-1,0) , \nonumber\\
&[S_1S_8^{(1)}S_{10}]^2 [S_4^{(3)}S_7^{(1)}S_{12}]:\ (0,0,1)
,\nonumber\\
& [S_1S_8^{(1)}S_{10}] [S_4^{(3)}S_7^{(1)}S_{12}]^2:\ (0,0,-1) ,
\nonumber
\end{align}
where the CS $H$-momenta are shown.  $S_{4}^{(1)}$, $S_4^{(3)}$
denote $S_4$ states with $(N^L)_j=2_{\bar{1}}, 2_{3}$,
respectively. Similarly, $S_{7,8}^{(1)}$, $S_{7,8}^{(3)}$ are
$S_{7,8}$ with $(N^L)_j=1_{\bar{1}}, 1_{3}$.  For oscillating
numbers $(N^L)_j$ of massless states, refer to the tables in
Appendix A.   CS' in Eq. (\ref{CSHmom}) are neutral under all the
gauge symmetries in this model, and fulfill the space group
selection rules of Eqs. (\ref{modinvk}) and (\ref{modinva}).
Hence, multiplication of the above CS' to an operator change only
the $H$-momentum vector by integers. Their VEVs are assumed to be
of the string scale on a vacuum.

\begin{table}[t]
\begin{center}
\begin{tabular}{|c|c|c|c|c|c|c|}
\hline  Visible states & SM notation& $B-L$ & $X$& $\Gamma$ &
$\Gamma'$&Label
\\&&&&&\\[-1.4em]
\hline $(0~0~0;0~0;1~0~-1)(0^8)'$ & ${\bf 1}_{\bf 0}(U_2)$& 0 &0 &
+2& 0&$S_0$
\\
 $(0^5;\frac{-2}{3}\frac{-2}{3}\frac{-1}{3})(\frac12\frac{-1}{2}~0;0^5)'$
 & ${\bf 1_0}(T_2^0)$& 0&0 & $+2$& 0&$S_{1}$\\
$(0^5;\frac{-2}{3}\frac{1}{3}\frac{2}{3})(\frac{-1}{2}\frac{1}{2}~0;0^5)'$
&  ${\bf 1_0}(T_2^0)$&0 &0 & $-2$& 0&$S_{2}$\\
$(0^5;\frac{1}{3}\frac{-2}{3}\frac{2}{3})(\frac{-1}{2}\frac{1}{2}~0;0^5)'$
&  ${\bf 1_0}(T_2^0)$&0 & 0& $0$& 0&$S_{3}$\\
$(0^5;\frac{1}{3}\frac{1}{3}\frac{-1}{3})(\frac{1}{2}\frac{-1}{2}~0;0^5)'$
&  $2\cdot{\bf 1_0}(T_2^0)$&0 & 0& $0$& 0&$S_{4}$\\
$(0^5;\frac{1}{3}\frac{1}{3}\frac{-1}{3})(\frac{-1}{2}\frac{1}{2}~0;0^5)'$
& $2\cdot{\bf 1_0}(T_2^0)$&0 & 0& $0$& 0&$S_{5}$\\
$(0^5;\frac{2}{3}\frac{2}{3}\frac{-2}{3})(0^8)'$ & $2\cdot{\bf
1_0}(T_4^0)$&0 &0& $0$& 0&$S_{6}$\\
$(0^5;\frac{-1}{3}\frac{-1}{3}\frac{-2}{3})(0^8)'$ & $7\cdot {\bf
1_0}(T_4^0)$&0 & 0& $+2$& 0&$S_{7}$\\
$(0^5;\frac{-1}{3}\frac{2}{3}\frac{1}{3})(0^8)'$ & $6\cdot{\bf
1_0}(T_4^0)$&0 & 0& $-2$& 0& $S_{8}$\\
$(0^5;\frac{2}{3}\frac{-1}{3}\frac{1}{3})(0^8)'$ & $6\cdot{\bf
1_0}(T_4^0)$&0 & 0& $0$& 0& $S_{9}$\\
$(0^5;1~0~0)(\frac{-1}{2}\frac{1}{2}~0;0^5)'$ &  $2\cdot{\bf
1_0}(T_6)$&0 & 0& $0$& 0&$S_{10}$\\
$(0^5;-1~0~0)(\frac{1}{2}\frac{-1}{2}~0;0^5)'$ & $2\cdot{\bf
1_0}(T_6)$&0 & 0& $0$& 0&$S_{11}$\\
 $(0^5;0~0~1)(\frac{-1}{2}\frac{1}{2}~0;0^5)'$
 & $2\cdot{\bf 1_0}(T_6)$&0 &0 &$-2$& 0&$S_{12}$\\
 $(0^5;0~0~-1)(\frac{1}{2}\frac{-1}{2}~0;0^5)'$
 & $2\cdot{\bf 1_0}(T_6)$&0 &0 & $+2$& 0&$S_{13}$\\
$(\frac{1}{4}\frac{1}{4}\frac{1}{4}\frac{1}{4}\frac{1}{4};
\frac{5}{12}\frac{5}{12}\frac{1}{12}
 )(\frac{1}{4}\frac{3}{4}~0;0^5)'$ & ${\bf 1_0}(T_1^0)$&$\frac12$ &$-\frac52$&
  $0$& \fbox{$+1$}&$S_{14}$\\
 $(\frac{1}{4}\frac{1}{4}\frac{1}{4}\frac{1}{4}\frac{1}{4};
\frac{5}{12}\frac{5}{12}\frac{1}{12}
 )(\frac{-3}{4}\frac{-1}{4}~0;0^5)'$ & ${\bf 1_0}(T_1^0)$& $\frac12$ &$-\frac52$&
  $\fbox{$-1$}$& 0&$S_{15}$\\ [0.1em]
$(\frac{-1}{4}\frac{-1}{4}\frac{-1}{4}\frac{-1}{4}\frac{-1}{4};
\frac{-1}{12}\frac{-1}{12}\frac{-5}{12}
 )(\frac{1}{4}\frac{3}{4}~0;0^5)'$ & ${\bf 1_0}(T_1^0)$& $-\frac12$ &$\frac52$&
 $\fbox{+1}$& 0&$S_{16}$\\
$(\frac{-1}{4}\frac{-1}{4}\frac{-1}{4}\frac{-1}{4}\frac{-1}{4};
\frac{-1}{12}\frac{-1}{12}\frac{-5}{12}
 )(\frac{-3}{4}\frac{-1}{4}~0;0^5)'$ & ${\bf 1_0}(T_1^0)$&$-\frac12$  &$\frac52$&
 $0$& \fbox{$-1$}&$S_{17}$\\
$(\frac{1}{4}\frac{1}{4}\frac{1}{4}\frac{1}{4}\frac{1}{4};
\frac{-7}{12}\frac{5}{12}\frac{1}{12}
 )(\frac{-1}{4}\frac{-3}{4}~0;0^5)'$
& ${\bf 1_0}(T_7^0)$&$\frac12$ &$-\frac52$& $\fbox{$-1$}$ & 0
&$S_{18}$\\
$(\frac{1}{4}\frac{1}{4}\frac{1}{4}\frac{1}{4}\frac{1}{4};
\frac{-7}{12}\frac{5}{12}\frac{1}{12}
 )(\frac{3}{4}\frac{1}{4}~0;0^5)'$ & ${\bf 1_0}(T_7^0)$&$\frac12$ &$-\frac52$&
  $0$& \fbox{$+1$}&$S_{19}$\\
$(\frac{1}{4}\frac{1}{4}\frac{1}{4}\frac{1}{4}\frac{1}{4};
\frac{5}{12}\frac{-7}{12}\frac{1}{12}
 )(\frac{-1}{4}\frac{-3}{4}~0;0^5)'$
& ${\bf 1_0}(T_7^0)$&$\frac12$ &$-\frac52$& $\fbox{+1}$& 0
&$S_{20}$\\
$(\frac{1}{4}\frac{1}{4}\frac{1}{4}\frac{1}{4}\frac{1}{4};
\frac{5}{12}\frac{-7}{12}\frac{1}{12}
 )(\frac{3}{4}\frac{1}{4}~0;0^5)'$ &
${\bf 1_0}(T_7^0)$&$\frac12$ &$-\frac52$& $+2$& \fbox{$+1$}
&$S_{21}$\\ &&&&&\\[-1.2em]
 $(\frac{-1}{4}\frac{-1}{4}\frac{-1}{4}\frac{-1}{4}\frac{-1}{4};
\frac{-1}{12}\frac{-1}{12}\frac{7}{12}
 )(\frac{-1}{4}\frac{-3}{4}~0;0^5)'$ &
${\bf 1_0}(T_7^0)$&$-\frac12$ &$\frac52$& $-2$& \fbox{$-1$}
&$S_{22}$\\
 $(\frac{-1}{4}\frac{-1}{4}\frac{-1}{4}\frac{-1}{4}\frac{-1}{4};
\frac{-1}{12}\frac{-1}{12}\frac{7}{12}
 )(\frac{3}{4}\frac{1}{4}~0;0^5)'$ &
${\bf 1_0}(T_7^0)$&$-\frac12$ &$\frac52$& $\fbox{$-1$}$& 0
&$S_{23}$\\
 $(\frac{-1}{4}\frac{-1}{4}\frac{-1}{4}\frac{-1}{4}\frac{-1}{4};
\frac{-3}{4}\frac{1}{4}\frac{1}{4}
 )(\frac{-1}{4}\frac{-3}{4}~0;0^5)'$ &
${\bf 1_0}(T_3)$&$-\frac12$ &$\frac52$& $-2$& \fbox{$-1$}
&$S_{24}$\\  &&&&&\\[-1.2em]
$(\frac{1}{4}\frac{1}{4}\frac{1}{4}\frac{1}{4}\frac{1}{4};
\frac{3}{4}\frac{-1}{4}\frac{-1}{4}
 )(\frac{1}{4}\frac{3}{4}~0;0^5)'$ &
${\bf 1_0}(T_9)$&$\frac12$ &$-\frac52$& $+2$ & \fbox{$+1$}
&$S_{25}$\\  &&&&&\\[-1.2em]
$(\frac{-1}{4}\frac{-1}{4}\frac{-1}{4}\frac{-1}{4}\frac{-1}{4};
\frac{1}{4}\frac{1}{4}\frac{-3}{4}
 )(\frac{-1}{4}\frac{-3}{4}~0;0^5)'$ & $2\cdot{\bf 1_0}(T_3)$&$-\frac12$ &$\frac52$
& $0$& \fbox{$-1$}&$S_{26}$\\  &&&&&\\[-1.2em]
  $(\frac{1}{4}\frac{1}{4}\frac{1}{4}\frac{1}{4}\frac{1}{4};
\frac{-1}{4}\frac{-1}{4}\frac{3}{4}
 )(\frac{1}{4}\frac{3}{4}~0;0^5)'$ &
${\bf 1_0}(T_9)$& $\frac12$&$-\frac52$& $0$& \fbox{$+1$}
&$S_{27}$\\  &&&&&\\[-1.2em]
$(\frac{1}{4}\frac{1}{4}\frac{1}{4}\frac{1}{4}\frac{1}{4};
\frac{-1}{4}\frac{-1}{4}\frac{-1}{4}
 )(\frac{3}{4}\frac{1}{4}~0;0^5)'$
& ${\bf 1_0}(T_3)$&$\frac12$ &$-\frac52$& $+2$ & \fbox{$+1$}
 &$S_{28}$ \\  &&&&&\\[-1.2em]
 $(\frac{-1}{4}\frac{-1}{4}\frac{-1}{4}\frac{-1}{4}
  \frac{-1}{4};\frac{1}{4}\frac{1}{4}\frac{1}{4})
  (\frac{-3}{4}\frac{-1}{4}~0;0^5)'$ &
$2\cdot{\bf 1_0}(T_9)$&$-\frac12$ &$\frac52$& $-2$& \fbox{$-1$}&
$S_{29}$
\\[0.2em]
\hline
\end{tabular}
\end{center}
\caption{Left-handed electromagnetically neutral $SO(10)'$
singlets. There is only one  untwisted sector singlet $S_0$. To
have a definition of parity, $S_{15}$, $S_{16}$, $S_{18}$,
$S_{20}$, and $S_{23}$ should not develop
VEVs.}\label{tb:Singneutral}
\end{table}

Then, on the vacuum with VEVs for $S_1$, $S_{4,7,8}^{(1,3)}$,
$S_{10}$, and $S_{12}$, the $H$-momentum conservation Eq.
(\ref{Hconsv}) can reduce to
\begin{eqnarray} \label{simpleHmom}
\sum_zR_j(z)\Longrightarrow {\rm integer} , ~~~j=1,~2,~3,
\end{eqnarray}
with the understanding that arbitrary number of CS' with ${\cal
O}(M_{\rm string})$ VEVs can be attached. Thus, if an operator's
$H$-momentum is an integer vector, proper CS' can be multiplied
such that the resultant $H$-momentum becomes $(-1, 1, 1)$ mod
$(12, 3, 12)$. Note that operators multiplied by (higher power of)
the above CS' are not suppressed, because the VEVs in Eq.
(\ref{CSHmom}) is assumed to be of order $M_{\rm string}$.
Moreover, $(N^L)_j$'s contributions to $H$-momentum also can be
always compensated by proper CS', because they just add integers
to $H_{\rm mom,0}$ as seen in Eq. (\ref{Hmom}).

$H$-momentum in $T_k$ sector is generally given by $(H_{\rm
mom,0}~{\rm in}~T_k)=(H_{\rm mom,0}~{\rm in}~T_1)\times k+({\rm
an~integer~vector})$. Accordingly the condition Eq.
(\ref{modinvk}) is equivalent to Eq. (\ref{simpleHmom}).  From now
on, we will require only (a) and (c) for Yukawa couplings with the
understanding proper CS' are multiplied.

\subsubsection{Phenomenologically desirable vacuum}
The phenomenologically desirable SSM vacuum is chosen by assigning
nonzero VEVs to {\it some} SM singlet fields such that

\noindent $\bullet$ unwanted exotics achieve heavy enough masses,

\noindent $\bullet$ U(1) gauge symmetries that are not observed at
low energies are broken, and

\noindent $\bullet$ $R$-parity violating couplings inducing too
rapid proton decay are sufficiently suppressed.

\noindent All the neutral singlets appearing in this model are
listed in Table \ref{tb:Singneutral}.

To attain the aims mentioned above, let us {\it choose a vacuum},
as one possibility, on which the following neutral singlets get
vanishing or non-vanishing VEVs:
\begin{align} \label{VEVsinglets}
& \langle S_0\rangle\ne 0 ,~ \langle S_1\rangle\ne 0 ,~ \cdots,
~\langle S_{13}\rangle\ne 0  , ~\langle S_{15}\rangle\ne 0 ,
~\langle S_{23}\rangle\ne 0  ,
~\langle S_{29} \rangle\ne 0 \\
& \langle S_{14}\rangle= \langle S_{16}\rangle= \langle
S_{17}\rangle=\cdots = \langle S_{22}\rangle = \langle
S_{24}\rangle= \langle S_{25}\rangle=\cdots = \langle S_{28}
\rangle=0.   \nonumber
\end{align}
In Sec. \ref{sec:Vecexotics}, we will show that the non-vanishing
VEVs in Eqs. (\ref{VEVsinglets}) are enough to give heavy masses
to all the exotics present in this model.

The VEVs of the singlets in Eq. (\ref{VEVsinglets}) break U(1)
symmetries in Eq. (\ref{gaugesym}) except U(1)$_Y$ and U(1)$_6$,
since all the neutral singlets don't carry the charges of U(1)$_Y$
and U(1)$_6$. The U(1)$_6$ generator is defined as
$Q_6=(0^8)(0,0,2;0^5)'$. In fact, all $Q_6$ nonzero charges are
carried only by the exotics as shown in Tables
\ref{tb:untwistedvi}--\ref{tb:T5}. All the observable matter
fields are neutral under U(1)$_6$.  Thus, in addition to photon
there exists another strictly massless U(1)$_6$ gauge boson which
is named as {\it exotic photon} ({\it exphoton} for abbreviation).
Since it couples only to superheavy exotic matter, the presence of
the ``exphoton'' is phenomenologically acceptable.

In Tables \ref{tb:standardch} and \ref{tb:Singneutral}, we
displayed the U(1)$_\Gamma$ and U(1)$_{\Gamma'}$ quantum numbers.
The U(1)$_\Gamma$ and U(1)$_{\Gamma'}$ are linear combinations of
U(1)s observed in this model. Their generators are defined as
\begin{eqnarray}
&&\Gamma=\textstyle X -(Q_2+Q_3)+\frac14(Q_4+Q_5) +6(B-L) ,
\label{U1Gamma}
\\
&&\Gamma'= \textstyle X +\frac14(Q_4+Q_5) +6(B-L) , \label{Gamma}
\end{eqnarray}
where
\begin{equation}
\begin{array}{l}
Q_2=(0^5;0,2,0)(0^8)',\quad  Q_3=(0^5;0,0,2)(0^8)'\\
 Q_4=(0^8)(2,0,0;0^5)',\quad Q_5=(0^8)(0,2,0;0^5)'\\
X=(-2,-2,-2,-2,-2;0^3)(0^8)' \\
 B-L=(\frac23,\frac23,\frac23,0^5)(0^8)'
 \end{array}
\end{equation}
$Q_4$ and $Q_5$ depend only on the hidden E$_8'$. The U(1)$_\Gamma$
symmetry will be used in Sec. \ref{sec:Rparity} for a discussion on
$R$-parity. We put boxes for $\Gamma^{(')}=\pm 1$ singlet fields. A
desirable vacuum toward an exact $R$-parity might be the one with
vanishing VEVs for all these boxed singlets. If the $R$-parity is
not exact, it should be an approximate symmetry valid at low energy
processes. These conditions should, of course, be consistent with
other phenomenological requirements such as large (small) enough
exotic mass terms ($\mu$ term).
In Table \ref{tb:standardch}, we also boxed some $D$ and
$\overline{D}$ fields which have different type $U(1)_\Gamma$
quantum numbers from those of $d$ and $d^c$ quarks. Namely, if the
parity defined from $U(1)_\Gamma$ is exact, these $D$ and
$\overline{D}$ do not mix with light quarks $d$ and $d^c$.

Note that the neutral singlets developing VEVs in Eq.
(\ref{VEVsinglets}) carry only zero or negative $\Gamma'$ charges:
$\Gamma'=0$ or $-1$. In fact, $\langle S_{29}\rangle$ break the
parity, however, in Sec. \ref{sec:Rparity} we will also show that
the light fields can still have a useful approximate $R$-parity.

\subsubsection{The third family in the untwisted sector}

Sixteen chiral fields in Eq. (\ref{eq:matter}) form a family. One
family appears in the untwisted sectors, $U_1$ and $U_3$. ${\rm
SU(2)_L}$ doublets are in $U_1$ and ${\rm SU(2)_L}$ singlets are
in $U_3$. The remaining two families arise from $T_4^0$. Since the
third family quarks are unique in being heavy, we assign the third
family to the untwisted sector fields. Indeed, there can exist
cubic couplings for the untwisted sector family by the coupling
$U_1U_2U_3$ allowed by the original selection rules (a), (b), and
(c). For this to be a viable interpretation, $H_u$ and $H_d$ in
$U_2$ must survive down to the electroweak scale.

\subsubsection{Light families and mixing angles}

With the VEVs of Eq. (\ref{VEVsinglets}), the (reduced) selection
rules allow also the mass terms of the first two families of the SSM
chiral matter. For example, $Q$ and $d^c$ in the $T_4^0$ sector can
couple together with $S_7$ or $S_1S_5$, if the oscillating number
carried by $S_7$ or $S_1S_5$ is compensated by a proper CS.  The
cross terms, $Q(U_1)$-$d^c(T_4^0)$ and $Q(T_4^0)$-$d^c(U_3)$ are
also possible through $S_7^2\cdot$CS (or $[S_1S_5]^2\cdot$CS). Thus
the $d^c-d$ mass matrix, $M^{(d)}$ takes the form
\begin{align}
&\quad Q(T_4^0)\ Q(T_4^0)\ Q(U_1)\nonumber\\
 \begin{array}{c}
d^c(T_4^0) \\ d^c(T_4^0) \\ d^c(U_3)
\end{array}
& \left(\begin{array}{ccc} a & b & x^{(d)}\\
\quad b\quad &\quad a\quad &\quad x^{(d)}\quad\\
x^{(d)} & x^{(d)} & 1 \end{array}
 \right)\langle H_d(U_2)\rangle  , \nonumber
\end{align}
where $a=S_7$ (or $S_1S_5$) and $x^{(d)}=S_7^2$ (or $[S_1S_5]^2$).
Here we set $\langle {\rm CS}\rangle=1$. The down-type quark mass
matrix is symmetric. For flavor democratic $T_4^0$ couplings, we
have a common entry $a$ instead of $a,b$ in the $2\times 2$
sub-matrix. But a flavor democratic form is one specific
representation of the ${\bf S}_2$ permutation symmetry. For a
general ${\bf S}_2$ representation for $T_4^0$ sector fields, the
upper left $2\times 2$ sub-matrix is of the form given above. So,
in general its determinant is nonzero. To have nonzero mixing
angles, the up-type quark mass matrix, $M^{(u)}$, should not align
to the down-type quark mass matrix,  $M^{(d)}$. The up-type
$u^c-u$ quark mass matrix is
\begin{align}
&\quad Q(T_4^0)\ Q(T_4^0)\ Q(U_1)\nonumber\\
 \begin{array}{c}
u^c(T_4^0) \\ u^c(T_4^0) \\ u^c(U_3)
\end{array}
& \left(\begin{array}{ccc} a' & b' & x^{(u)}\\
\quad b'\quad &\quad a'\quad &\quad x^{(u)}\quad\\
y^{(u)} & y^{(u)} & 1 \end{array}
 \right)\langle H_u(U_2)\rangle  , \nonumber
\end{align}
where
\begin{align}
\{a', b'\}=\{S_9, ~S_3S_4 \},\ \  x^{(u)}=\{S_7S_9, ~S_1S_5S_9,
~S_7S_3S_4\},\ \ y^{(u)}=\{S_8S_9, ~S_2S_4S_9, ~S_8S_3S_4 \} .
\nonumber
\end{align}
$a'$ and $b'$ which are linear combinations of $S_9$ and $S_3S_4$
can be different in principle. In $M^{(u)}$, proper CS
multiplications are assumed. Unlike $M^{(d)}$, $M^{(u)}$ is not
symmetric.

Similarly, the charged lepton mass matrix $M^{(e)}$ is
\begin{align}
&\quad L(T_4^0)\quad L(T_4^0)\quad L(U_1)\nonumber\\
 \begin{array}{c}
e^c(T_4^0) \\ e^c(T_4^0) \\ e^c(U_3)
\end{array}
& \left(\begin{array}{ccc} a'' & b'' & x^{(e)}\\
\quad b''\quad &\quad a''\quad &\quad x^{(e)}\quad\\
 y^{(e)} &  y^{(e)} & 1 \end{array}
 \right)\langle H_d(U_2)\rangle  , \nonumber
\end{align}
where
\begin{align}
&\{a'', b''\}=\{S_7, ~S_1S_5\},\ \ x^{(e)}=\{S_7S_8, ~S_1S_5S_8,
~S_7S_2S_4\},\ \ y^{(e)}=\{S_7S_9, ~S_1S_5S_9,
~S_7S_3S_4\}.\nonumber
\end{align}

Neutrinos obtain mass. With the following Dirac and Majorana mass
terms, the seesaw type light neutrino masses are possible:
\begin{align}
&\quad L(T_4^0)\quad L(T_4^0)\quad L(U_1)\nonumber\\
{\rm Dirac}:\
 \begin{array}{c}
\nu^c(T_4^0) \\ \nu^c(T_4^0) \\ \nu^c(U_3)
\end{array}
& \left(\begin{array}{ccc} c & c & x^{(\nu)}\\
\quad c\quad &\quad c\quad &\quad x^{(\nu)}\quad\\
 y^{(\nu)} &  y^{(\nu)} & 1 \end{array}
 \right)\langle H_u(U_2)\rangle  , \nonumber\\
 &\nonumber\\
&\quad \nu^c(T_4^0)\quad \nu^c(T_4^0)\quad \nu^c(U_3)\nonumber\\
{\rm Majorana}:\
 \begin{array}{c}
\nu^c(T_4^0) \\ \nu^c(T_4^0) \\ \nu^c(U_3)
\end{array}
& \left(\begin{array}{ccc} M_2 & M_2 & M_1\\
\quad M_2\quad &\quad M_2\quad &\quad M_1\quad\\
 M_1 &  M_1 & M_0 \end{array}
 \right)\nonumber
\end{align}
where
\begin{align}
&c=\{S_9, ~S_3S_4\}\ \ x^{(\nu)}=\{S_8S_9, ~S_2S_4S_9,
~S_8S_3S_4\},\ \ y^{(\nu)}=\{S_7S_9, S_1S_5S_9, ~S_7S_3S_4\} ,
\nonumber
\end{align}
and
\begin{align}
&M_0=[S_{23}S_{29}]^2[S_7]^4,\ \ M_1=[S_{23}S_{29}]^2[S_7]^3,\ \
M_2=[S_{23}S_{29}]^2[S_7]^2.\nonumber
\end{align}
Therefore, the vacuum (\ref{VEVsinglets}) can give successful
quark and lepton mass matrices.

\subsubsection{Higgs doublets and $\mu$ term} \label{higgsdoublet}

Vectorlike electroweak doublet fields, $H_u(Y=\frac12)$ and
$H_d(Y=-\frac12)$, appear in $U_2$, $T_4^0$, $T_6$, $T_3,$ and
$T_9$.
The selection rules (b) and (c) in Sec. (\ref{rules}) allow
interactions of $U_2U_2\times$CS and $U_2U_2T_6T_6\times$CS. Among
these interactions, $[H_u(U_2)H_d(U_2)]\times (S_0\cdot {\rm
CS}+S_{10}S_{13}\cdot {\rm CS})$ are present. We regard $\{H_u(U_2),
H_d(U_2)\}$ as the MSSM Higgs fields.  TeV scale VEV of $(S_0\cdot
{\rm CS}+S_{10}S_{13}\cdot {\rm CS})$ gives the MSSM ``$\mu$'' term.
We will discuss it again later.

The selection rules permit $T_6T_6\times$CS couplings. So,
$H_u(T_6)H_d(T_6)\times$CS couplings are present. Hence two pairs
of $H_u$ and $H_d$ from $T_6$ obtain heavy mass by string scale
VEV of CS.  The selection rules admit also $T_4^0T_4^0T_4^0$
couplings.  So there exist $H_u(T_4^0)H_d(T_4^0)S_6(T_4^0)$
couplings, from which two pairs of $H_u$ and $H_d$ in $T_4^0$ also
become heavy by string scale VEVs of $S_6$.\footnote{We ignore a
possible permutation symmetry at this level of study.} There
remains one $H_d(T_4^0)$ at this level.

$T_3T_9\times$CS couplings are also allowed. Thus, there exist
couplings of $H_u(T_9)H_d(T_3)\times$CS, and by a VEV of CS one pair
of $\{H_u(T_9), H_d(T_3)\}$ is made heavy. Thus, there remains one
$H_u(T_9)$ also at this level.

The remaining $H_d$ in $T_4^0$ and $H_u$ in $T_9$ can also be made
heavy via the coupling $[H_u(T_9)H_d(T_4^0)]\times \langle
S_4S_{29}\rangle$. This coupling is one of $T_9T_4^0T_2^0T_9$
interactions, which satisfies the selection rules. To study the
masses in more detail, we list the full $H_uH_d$ couplings in Table
\ref{tb:muterms}.

\begin{table}[h]
\begin{center}
\begin{tabular}{|c|c|}
\hline &\\[-1.2em] Pairs & ~Masses ($\times$proper CS)
\\[0.2em] \hline
$\{H_u(U_2),~ H_d(U_2)\}$ & $S_0$, $S_{10}S_{13}$
\\
$\{H_u(U_2),~ H_d(T_6)\}$ & $S_{10}$
\\
$\{H_u(U_2),~ H_d(T_4^0)\}$ & $S_{10}S_4$
\\
$\{H_u(U_2),~ H_d(T_3)\}$ & $0$
\\ \hline
$\{H_u(T_6),~ H_d(U_2)\}$ & $S_{13}$
\\
$\{H_u(T_6),~ H_d(T_6)\}$ & $1$
\\
$\{H_u(T_6),~ H_d(T_4^0)\}$ & $S_4$
\\
$\{H_u(T_6),~ H_d(T_3)\}$ & $0$
\\ \hline
$\{H_u(T_4^0),~ H_d(U_2)\}$ & $S_{13}S_5$
\\
$\{H_u(T_4^0),~ H_d(T_6)\}$ & $S_5$
\\
$\{H_u(T_4^0),~ H_d(T_4^0)\}$ & $S_6$
\\
$\{H_u(T_4^0),~ H_d(T_3)\}$ & $0$
\\ \hline
$\{H_u(T_9),~ H_d(U_2)\}$ & $S_{13}S_{29}$
\\
$\{H_u(T_9),~ H_d(T_6)\}$ & $S_{29}$
\\
$\{H_u(T_9),~ H_d(T_4^0)\}$ & $S_4S_{29}$
\\
$\{H_u(T_9),~ H_d(T_3)\}$ & $1$
\\[0.1em]
\hline
\end{tabular}
\end{center}
\caption{Mass terms for $H_u$ and $H_d$. CS are products of
singlet fields given in Eq. (2.19). Proper CS are assumed to be
multiplied such that the $H$-momentum becomes $(-1,1,1)$ mod
$(12,3,12)$. We set $\langle {\rm CS}\rangle=1$. For $\mu$
solution we assume that a modulus is involved in $S_0$ or
$S_{10}S_{13}$.}\label{tb:muterms}
\end{table}
Now we can represent a schematic form of the $7\times 7\ H_uH_d$
mass matrix as
\begin{eqnarray}
&&\quad\quad\ \ \ H_0 \  H_6 \ H_6 \ H_4 \ H_4 \ H_4 \ H_3\nonumber\\
&&\begin{array}{c} H^0\\ H^6\\ H^6\\ H^4\\ H^4\\ H^9\\
H^9\end{array}\left(
\begin{array}{ccccccc}
 \triangle \ & \star\ &\star\  &\star'\ &\star'\ &\star'\ &\ 0\ \\
 \ast \ &\times \ &\times \ &>  \ &> \ &> \ &0   \\
 \ast \ &\times \ &\times \ &>  \ &> \ &> \ &0   \\
 \ast' \ &< \ &< \ &\vee \ &\vee \ &\vee  \ &0   \\
 \ast' \ &< \ &< \ &\vee \ &\vee \ &\vee \ &0  \\
 \ast'' \ &\Diamond \ &\Diamond \ &\Diamond' \ &\Diamond'
 \ &\Diamond' \ &\times \\
 \ast'' \ &\Diamond \ &\Diamond \ &\Diamond' \ &\Diamond'
 \ &\Diamond' \ &\times
\end{array}\right) . \label{seesaw}
\end{eqnarray}
Here $H^0$, $H^6$, $H^4$, and $H^9$ indicate $ H_u(U_2)$,
$H_u(T_6)$, $H_u(T_4)$, and $H_u(T_9)$, respectively.  Similarly,
$H_0\equiv H_d(U_2)$, $H_4\equiv H_d(T_4^0)$, and $H_3\equiv
H_d(T_3)$.
$\triangle$ denotes non-vanishing VEVs by $S_0$ and $S_{10}S_{13}$,
$\triangle\equiv S_0+S_{10}S_{13}$. As mentioned earlier, we tacitly
assume proper VEVs of CS, which are of string scale, are multiplied
to fulfill the selection rule (b) discussed in Sec. \ref{rules}.
$\times$s stand for non-vanishing VEVs by CS. $\star$ and $\star'$
are VEVs of $S_{10}$ and $S_{10}S_4$, and $\ast$, $\ast'$, $\ast''$
are those of $S_{13}$, $S_{13}S_5$, $S_{13}S_{29}$.  $>$, $<$, and
$\vee$ correspond to VEVs of $S_4$, $S_5$, and $S_6$, respectively.
$\Diamond$ and $\Diamond'$ are VEVs of $S_{29}$ and $S_4S_{29}$.
Since any neutral singlets with non-vanishing VEVs do not carry
positive $\Gamma'$ charges, zero entries in the above matrix Eq.
(\ref{seesaw}), which are associated with $H_d(T_3)$, should be
exactly zeros.

We suppose relatively small VEVs for $S_{10}$ and $S_{13}$
compared to the other VEVs of neutral singlets:
\begin{eqnarray}
S_{10}, ~S_{13} ~\lesssim ~{\cal O}(M_{\rm string}) .
\end{eqnarray}
Then the mixing angle between $\{H_u(U_2), H_d(U_2)\}$ and the
other $H_u$-$H_d$ pairs is suppressed, and the effective ``$\mu$''
coefficient of $H_u(U_2)H_d(U_2)$ is estimated as
\begin{eqnarray}
\mu\sim S_0+{\cal O}(S_{10}S_{13}/M_{\rm string}) .
\end{eqnarray}
If one VEV among $S_0$, $S_{10}$, and $S_{13}$ is left
undetermined at the string scale, $\mu$ is also undetermined in
the SUSY limit. With soft SUSY breaking terms, however, $\mu$ (and
Higgs VEVs) could be fixed around TeV scale.  In the limit
$\mu\rightarrow 0$, an accidental Peccei-Quinn symmetry revives.
We do not discuss it in this paper.

\subsubsection{Vectorlike $D^{-1/3}$ and $\overline{D}^{1/3}$}

The \Qem$=\mp\frac13$ colored fields $D^{-1/3}$ and
$\overline{D}^{1/3}$ appear only in twisted sectors $T_6$, $T_4^0$
$T_3$, and $T_9$.  Three pairs of $\{D(T_6)$ and
$\overline{D}(T_6)\}$ can be removed from low energy field spectra
via $D(T_6)\overline{D}(T_6)\times$CS.

The coupling $D(T_4^0)\overline{D}(T_4^0)S_6(T_4^0)$ remove two
pairs of $D$ and $\overline{D}$ in $T_4^0$, leaving one
$\overline{D}$ in $T_4^0$.  The coupling of the form
$D(T_9)\overline{D}(T_3)\times$CS is present, and so one pair of
$D$ and $\overline{D}$ is removed at this level, leaving one $D$
in $T_9$.

The remaining $\overline{D}(T_4^0)$ and $D(T_9)$ can be heavy via
the two couplings $[D(T_9)\overline{D}(T6)]\times\langle
S_4S_{23}\cdot{\rm CS}\rangle$ and
$[D(T6)\overline{D}(T_4^0)]\times\langle S_5\cdot{\rm CS}\rangle$.
Note that here $D(T_6)$ and $\overline{D}(T_6)$ are already
coupled to each other to have the mass term with a VEV of CS.
Therefore it is obvious that all $\{D, \overline{D}\}$ obtain
masses.  We list all $D$-$\overline{D}$ couplings in Table
\ref{tb:DDbarmass}.

\begin{table}[h]
\begin{center}
\begin{tabular}{|c|c|}
\hline & \\[-1.2em] Pairs & ~Masses ($\times$proper CS)
\\[0.2em] \hline &\\[-1.2em]
$\{D(T_6),~\overline{D}(T_6) \}$ & $1$
\\
$\{D(T_6), ~\overline{D}(T_4)\}$ & $S_5$
\\
$\{D(T_6), ~\overline{D}(T_3)\}$ & $0$
\\ \hline
$\{D(T_4), ~\overline{D}(T_6)\}$ & $S_4$
\\
$\{D(T_4),~\overline{D}(T_4) \}$ & $S_6$
\\
$\{D(T_4),~\overline{D}(T_3) \}$ & $S_7$\fbox{$S_{15}$}
\\ &\\[-1.2em]\hline &\\[-1.2em]
$\{D(T_9),~\overline{D}(T_6)\}$ & \fbox{$S_{23}$}$S_4$
\\ &\\[-1.2em]
$\{D(T_9),~\overline{D}(T_4) \}$ & \fbox{$S_{23}$}$S_6$
\\
$\{D(T_9),~\overline{D}(T_3) \}$ & $1$
\\ \hline
\end{tabular}
\end{center}
\caption{Mass terms for $D$ and $\overline{D}$.
 CS are products of singlet fields
given in Eq. (2.19). Proper CS are assumed to be multiplied such
that the $H$-momentum becomes $(-1,1,1)$ mod $(12,3,12)$.  We set
$\langle {\rm CS}\rangle=1$.}\label{tb:DDbarmass}
\end{table}
The $8\times 8$ $D$-$\overline{D}$ mass matrix is of the form
\begin{eqnarray}
&&\quad\quad\  \ \ \overline{D}_6\ \overline{D}_6\
 \overline{D}_6\ \overline{D}_4\ \overline{D}_4\ \overline{D}_4\
 \overline{D}_3\nonumber\\
&&\begin{array}{c}   D_6\\ D_6\\ D_6\\ D_4\\ D_4\\ D_9\\ D_9\\
\end{array}\left(
\begin{array}{cccccccc}
\ \times \ &\times\  &\times  &\ <\ &\ <\ & <\ &  0 \  \\
\times &\times &\times &\ <\ &\ <\ & <\ &   0 \ \\
\times &\times &\times &\ <\ &\ <\ & <\ &   0 \ \\
 > & > & > &\vee  &\vee &\vee & \boxtimes \ \\
 > &\ > &\ > &\vee & \vee & \vee & \boxtimes \ \\
\ \square &\ \square &\ \square  &\ \square' & \square' & \square' &\times \\
\ \square &\ \square &\ \square  &\ \square' & \square' & \square'
&\times
\end{array}\right) , \label{DDbar}
\end{eqnarray}
where $D_6, \overline{D}_4$, etc. mean $D(T_6), \overline{D}(T_4)$,
etc. $\times$, $<$, $>$, and $\vee$ entries stand again for VEVs of
CS, $S_5$, $S_4$, and $S_6$, respectively. $\square$, $\square'$,
and $\boxtimes$ denote VEVs of $S_{23}S_4$, $S_{23}S_6$, and
$S_7S_{15}$. Through the mass terms in Eq. (\ref{DDbar}), all $D$s
and $\overline{D}$s are paired to be superheavy. Mixing terms
between $d^c$s in $U_3$, $T_4^0$ and $D$s in $T_6$, $T_4^0$, $T_9$
can not arise in any manner. It is because the negative $\Gamma'$
charges carried by such mixing terms cannot be compensated by
neutral singlets with non-zero VEVs.

This shows that the odd $\Gamma$ and \Qem$=-\frac13$ quarks of Table
\ref{tb:standardch} can mix among themselves, but not with $D(T_6)$,
$D(T_4^0),$ $\overline{D}(T_6)$ and $\overline{D}(T_4^0)$, in the
limit $S_{23}\to 0$. So, the down-type quarks have additional
contribution to the mass matrix by mixing with $D(T_9)$ and
$\overline{D}(T_3)$, and non-vanishing quark mixing is achieved in
general.

Even if $S_{15}=0$ (so $\boxtimes =0$), all $D$ and $\overline{D}$
still obtain masses because the determinant of Eq. (\ref{DDbar}) is
nonzero. If $S_{23}=0$ (so $\square=\square'=0$), however, the above
type mass mixing does not give a mass to one pair of
$D$-$\overline{D}$. Hence it seems necessary to have at least one
$\Gamma$ odd singlet obtain a VEV. Let us choose the VEV $\langle
S_{23}\rangle$ as the parameter contributing to $P$ violating terms
among the low energy fields.

\section{Vectorlike exotics}\label{sec:Vecexotics}

Among the phenomenological conditions, the exotics mass condition
must be satisfied at any cost. In this model, exotic fields
appears in the $T_1^{\pm}$, $T_2^{\pm}$, $T_4^{\pm}$, and
$T_7^{\pm}$ (or $T_5^{\pm}$) sectors.  The color triplet exotics
carry the electromagnetic charges of $0$, $\pm\frac13$. The
doublet and singlet exotics carry also fractional electromagnetic
charges: \Qem$=\pm\frac23,\mp\frac13$. Color exotics could form
color singlet states with fractional electromagnetic charges.
Searches for fractionally charged particles have not given any
positive evidence, and hence all exotics on the vacuum we choose
should be heavy enough.  Let us proceed to discuss how the
vectorlike exotic states achieve masses.

\subsection{Color exotics}\label{subsec:colex}

In Table (\ref{tb:Cexotics}), we list the color exotics found in
our model. They are singlets under ${\rm SU(2)_L}$. As seen in the
table, the color exotics are vectorlike under the SM gauge
symmetry.
\begin{table}[h]
\begin{center}
\begin{tabular}{|c|c|c|}
\hline &&\\ [-1em]
  Color exotics & SU(3)$_c$(Sector) & $({\alpha\ {\rm or\ }
  \overline{\alpha}})^{{Q}_{\rm em}}$
\\[0.2em]
\hline
 $\left(\underline{\frac{-7}{12},\frac{5}{12},\frac{5}{12}};
\frac{1}{12},\frac{1}{12};
\frac{-5}{12},\frac{-1}{12},\frac{3}{12}\right)
\left(\frac{3}{12},\frac{5}{12},\frac{-4}{12};0^5\right)'$ & ${\bf
3}(T_1^+)$ &  $\alpha_1^0$
\\[0.4em]

$\left(\underline{\frac{5}{6},\frac{-1}{6},\frac{-1}{6}};
\frac{1}{6},\frac{1}{6};
\frac{1}{6},\frac{-1}{6},\frac{-1}{2}\right)
\left(\frac{1}{2},\frac{-1}{6},\frac{1}{3};0^5\right)'$ & ${\bf
3}^*(T_2^+)$ &  $\overline{\alpha}_2^0$
\\[0.4em]
$\left(\underline{\frac{-5}{6},\frac{1}{6},\frac{1}{6}};
\frac{-1}{6},\frac{-1}{6};
\frac{-1}{2},\frac{-1}{6},\frac{-1}{6}\right)
\left(\frac{-1}{2},\frac{1}{6},\frac{-1}{3};0^5\right)'$ & ${\bf
3}(T_2^-)$ &  $\alpha_3^0$
\\[0.4em]
$\left(\underline{\frac{-5}{6},\frac{1}{6},\frac{1}{6}};
\frac{-1}{6},\frac{-1}{6};
\frac{-1}{6},\frac{1}{6},\frac{-1}{2}\right)
\left(0,\frac{-1}{3},\frac{-1}{3};0^5\right)'$ & ${\bf 3}(T_4^+)$
 & $2\cdot \alpha_4^0$
\\
$\left(\underline{\frac{5}{6},\frac{-1}{6},\frac{-1}{6}};
\frac{1}{6},\frac{1}{6};
\frac{-1}{2},\frac{1}{6},\frac{1}{6}\right)
\left(0,\frac{1}{3},\frac{1}{3};0^5\right)'$ &  ${\bf 3}^*(T_4^-)$
 &  $3\cdot \overline{\alpha}_5^0$
\\[0.4em]
$\left(\underline{\frac{2}{3},\frac{-1}{3},\frac{-1}{3}};
\frac{1}{3},\frac{1}{3}; \frac{1}{3},\frac{-1}{3},0\right)
\left(0,\frac{-1}{3},\frac{-1}{3};0^5\right)'$ & ${\bf
3}^*(T_4^+)$ &  $2\cdot \overline{\alpha}_6^{-1/3}$
\\[0.4em]
$\left(\underline{\frac{-2}{3},\frac{1}{3},\frac{1}{3}};
\frac{-1}{3},\frac{-1}{3}; 0,\frac{-1}{3},\frac{-1}{3}\right)
\left(0,\frac{1}{3},\frac{1}{3};0^5\right)'$ &  ${\bf 3}(T_4^-)$ &
$2\cdot \alpha_7^{1/3}$
\\[0.4em]
\hline
\end{tabular}
\end{center}
\caption{Color exotics of \Qem$=0,\pm \frac13$. Color {\bf 3} and
${\bf 3}^*$ with \Qem$=\pm\frac13$ are exotics.
}\label{tb:Cexotics}
\end{table}
They all can achieve masses when the neutral singlets in Eq.
(\ref{VEVsinglets}) get VEVs. To prove this, we don't have to study
the full mass matrix for the vectorlike exotics. Instead, we will
suggest just some couplings enough to show that they are heavy. In
Table \ref{tb:colorexmass}, we present the minimal number of
couplings yielding their masses.
\begin{table}[h]
\begin{center}
\begin{tabular}{|c|c|}
\hline &\\[-1.2em] Pairs & ~Masses ($\times$proper CS)
\\[0.2em] \hline
 $1\times\{\alpha_3^0(T_2^-),~ \overline{\alpha}_2^0(T_2^+)\}$
& $S_4S_{12}$
\\
$2\times\{\alpha_4^0(T_4^+),~ \overline{\alpha}_5^0(T_4^-)\}$ &
$S_9$, $S_3S_4$
\\
$1\times\{\alpha_1^0(T_1^+),~ \overline{\alpha}_5^0(T_4^-)\}$ &
$S_9S_{13}S_{29}$
\\
$2\times\{\alpha_7^{1/3}(T_4^-),~
\overline{\alpha}_6^{-1/3}(T_4^+)\}$ & $S_8$, $S_2S_4$
\\[0.2em]
\hline
\end{tabular}
\end{center}
\caption{Mass terms for color exotics. CS are products of singlet
fields given in Eq. (2.19). Proper CS are assumed to be multiplied
such that the $H$-momentum becomes $(-1,1,1)$ mod $(12,3,12)$. We
set $\langle {\rm CS}\rangle=1$.} \label{tb:colorexmass}
\end{table}
Since all the vectorlike exotics in Table \ref{tb:Cexotics} can
pair up with neutral singlets, they can be removed from low energy
field spectra.

\subsection{Doublet exotics}\label{subsec:dbex}

In this model there are ${\rm SU(2)_L}$ doublet fields carrying
exotic electromagnetic charges.  They are ${\rm SU(3)_c}$ singlets
but possess the charges of $Y=\pm\frac16$ (or $Q_{\rm
em}=\pm\frac23,~\mp\frac13$).  In Table \ref{tb:Dexotics}, all
doublet exotics are collected.
\begin{table}[h]
\begin{center}
\begin{tabular}{|c|c|c|}
\hline  &&\\ [-1em]
Doublet exotics  & $[{\rm
SU(2)_L}]^Y$(Sector)&Label
\\
\hline $\left(\frac{-1}{12},\frac{-1}{12},\frac{-1}{12};
\underline{\frac{7}{12},\frac{-5}{12}};
\frac{1}{12},\frac{5}{12},\frac{-3}{12}\right)
\left(\frac{3}{12},\frac{5}{12},\frac{-4}{12};0^5\right)'$  &
${\bf 2}^{-1/6}(T_1^+)$& $\overline{\delta}_1$
\\[0.4em]
 $\left(\frac{1}{12},\frac{1}{12},\frac{1}{12};
\underline{\frac{-7}{12},\frac{5}{12}};
\frac{3}{12},\frac{-1}{12},\frac{-1}{12}\right)
\left(\frac{-9}{12},\frac{1}{12},\frac{4}{12};0^5\right)'$  &
${\bf 2}^{1/6}(T_1^-)$& ${\delta}_2$
\\[0.4em]
$\left(\frac{1}{12},\frac{1}{12},\frac{1}{12};
\underline{\frac{-7}{12},\frac{5}{12}};
\frac{3}{12},\frac{-1}{12},\frac{-1}{12}\right)
\left(\frac{3}{12},\frac{1}{12},\frac{-8}{12};0^5\right)'$  &
${\bf 2}^{1/6}(T_1^-)$& ${\delta}_3$
\\[0.4em]
$\left(\frac{1}{6},\frac{1}{6},\frac{1}{6};
\underline{\frac{5}{6},\frac{-1}{6}};
\frac{-1}{6},\frac{1}{6},\frac{-1}{2}\right)
\left(0,\frac{-1}{3},\frac{-1}{3};0^5\right)'$  & $3\cdot {\bf
2}^{-1/6}(T_4^+)$& $\overline{\delta}_4$
\\[0.4em]
$\left(\frac{-1}{3},\frac{-1}{3},\frac{-1}{3};
\underline{\frac{-2}{3},\frac{1}{3}};
\frac{1}{3},\frac{-1}{3},0\right)
\left(0,\frac{-1}{3},\frac{-1}{3};0^5\right)'$ &  $2\cdot {\bf
2}^{-1/6}(T_4^+)$& $\overline{\delta}_5$
\\[0.4em]
$\left(\frac{-1}{6},\frac{-1}{6},\frac{-1}{6};
\underline{\frac{-5}{6},\frac{1}{6}};
\frac{-1}{2},\frac{1}{6},\frac{1}{6}\right)
\left(0,\frac{1}{3},\frac{1}{3};0^5\right)'$ &  $2\cdot {\bf
2}^{1/6}(T_4^-)$& ${\delta}_6$
\\[0.4em]
$\left(\frac{1}{3},\frac{1}{3},\frac{1}{3};
\underline{\frac{2}{3},\frac{-1}{3}};
0,\frac{-1}{3},\frac{-1}{3}\right)
\left(0,\frac{1}{3},\frac{1}{3};0^5\right)'$   & $2\cdot {\bf
2}^{1/6}(T_4^-)$& ${\delta}_7$
\\[0.4em]
$\left(\frac{-1}{12},\frac{-1}{12},\frac{-1}{12};
\underline{\frac{7}{12},\frac{-5}{12}};
\frac{1}{12},\frac{-7}{12},\frac{-3}{12}\right)
\left(\frac{-3}{12},\frac{-1}{12},\frac{8}{12};0^5\right)'$ &
${\bf 2}^{-1/6}(T_7^+)$&$\overline{\delta}_8$
\\[0.4em]
  $\left(\frac{1}{12},\frac{1}{12},\frac{1}{12};
\underline{\frac{-7}{12},\frac{5}{12}};
\frac{-9}{12},\frac{-1}{12},\frac{-1}{12}\right)
\left(\frac{-3}{12},\frac{-5}{12},\frac{4}{12};0^5\right)'$ &
${\bf 2}^{1/6}(T_7^-)$&${\delta}_9$
\\[0.4em]

\hline
\end{tabular}
\end{center}
\caption{SU(2) doublet exotics with \Qem$=\pm\frac23,\mp
\frac13$.}\label{tb:Dexotics}
\end{table}
All the vectorlike doublet exotics in Table \ref{tb:Dexotics}
could achieve masses via couplings with neutral singlets
developing VEVs. The minimal number of couplings for them to be
heavy are displayed in Table \ref{tb:doubexmass}.
\begin{table}[h]
\begin{center}
\begin{tabular}{|c|c|}
\hline &\\[-1.2em] Pairs & ~Masses ($\times$proper CS)
\\[0.2em] \hline &\\[-1.2em]
 $1\times\{\overline{\delta}_4(T_4^+),~ \delta_2(T_1^-)\}$ &
\fbox{$S_{23}$}
\\
$2\times\{\overline{\delta}_5(T_4^+),~ \delta_7(T_4^-)\}$ & $S_8$,
$S_2S_4$
\\
$1\times\{\overline{\delta}_1(T_1^+),~ \delta_9(T_7^-)\}$ & $S_9$,
$S_3S_4$
\\
$1\times\{\overline{\delta_8}(T_7^+),~ \delta_3(T_1^-)\}$ & $S_8$,
$S_2S_4$
\\
$2\times\{\overline{\delta}_4(T_4^+),~ \delta_6(T_4^-)\}$ & $S_9$,
$S_3S_4$
\\
[0.2em] \hline
\end{tabular}
\end{center}
\caption{Mass terms for doublet exotics. CS are products of
singlet fields given in Eq. (2.19). Proper CS are assumed to be
multiplied such that the $H$-momentum becomes $(-1,1,1)$ mod
$(12,3,12)$. We set $\langle {\rm CS}\rangle=1$.}
\label{tb:doubexmass}
\end{table}
Hence, all the doublet exotics can obtain masses.

\subsection{Singlet exotics}\label{subsec:singex}

There are 38 kinds (in terms of gauge quantum numbers) of singlet
exotics, as collected in Tables \ref{tb:Sexotics}.
\begin{table}[t]
{\small
\begin{center}
\begin{tabular}{|c|c|c|}
\hline &&\\[-1em]
 States & SM notation & Label
\\&&\\[-1em]
\hline  &&\\[-1.1em]
$\left(\frac{1}{3},\frac{1}{3},\frac{1}{3};\frac{-1}{3},\frac{-1}{3};
\frac{-1}{3},\frac{1}{3},0\right)
\left(\frac{-1}{2},\frac{-1}{6},\frac{-2}{3};0^5\right)'$ & ${\bf
1}_{2/3}(T_2^+)$
%
& $\xi_{1}$
\\
$\left(\frac{1}{3},\frac{1}{3},\frac{1}{3};\frac{-1}{3},\frac{-1}{3};
\frac{-1}{3},\frac{1}{3},0\right)
\left(\frac{1}{2},\frac{-1}{6},\frac{1}{3};0^5\right)'$ & ${\bf
1}_{2/3}(T_2^+)$
%
& $\xi_{2}$
\\
$\left(\frac{-1}{6},\frac{-1}{6},\frac{-1}{6};\frac{1}{6},\frac{1}{6};
\frac{-5}{6},\frac{-1}{6},\frac{-1}{2}\right)
\left(\frac{1}{2},\frac{-1}{6},\frac{1}{3};0^5\right)'$ & ${\bf
1}_{-1/3}(T_2^+)$
%
& $\eta_{1}$
\\
$\left(\frac{-1}{6},\frac{-1}{6},\frac{-1}{6};\frac{1}{6},\frac{1}{6};
\frac{1}{6},\frac{-1}{6},\frac{1}{2}\right)
\left(\frac{-1}{2},\frac{5}{6},\frac{1}{3};0^5\right)'$ & ${\bf
1}_{-1/3}(T_2^+)$
%
& $\eta_{2}$
\\
$\left(\frac{-1}{6},\frac{-1}{6},\frac{-1}{6};\frac{1}{6},\frac{1}{6};
\frac{1}{6},\frac{-1}{6},\frac{1}{2}\right)
\left(\frac{-1}{2},\frac{-1}{6},\frac{-2}{3};0^5\right)'$ & ${\bf
1}_{-1/3}(T_2^+)$
%
& $\eta_{3}$
\\
$\left(\frac{-1}{6},\frac{-1}{6},\frac{-1}{6};\frac{1}{6},\frac{1}{6};
\frac{1}{6},\frac{-1}{6},\frac{1}{2}\right)
\left(\frac{1}{2},\frac{-1}{6},\frac{1}{3};0^5\right)'$ &
$2\cdot{\bf 1}_{-1/3}(T_2^+)$
%
& $\eta_{4}, \eta_{5}$
\\
$\left(\frac{-1}{3},\frac{-1}{3},\frac{-1}{3};\frac{1}{3},\frac{1}{3};
0,\frac{1}{3},\frac{1}{3}\right)
\left(\frac{-1}{2},\frac{1}{6},\frac{-1}{3};0^5\right)'$ & ${\bf
1}_{-2/3}(T_2^-)$
%
& $\overline{\xi}_{3}$
\\
$\left(\frac{1}{6},\frac{1}{6},\frac{1}{6};\frac{-1}{6},\frac{-1}{6};
\frac{-1}{2},\frac{-1}{6},\frac{5}{6}\right)
\left(\frac{-1}{2},\frac{1}{6},\frac{-1}{3};0^5\right)'$ & ${\bf
1}_{1/3}(T_2^-)$
%
& $\overline{\eta}_{6}$
\\
$\left(\frac{1}{6},\frac{1}{6},\frac{1}{6};\frac{-1}{6},\frac{-1}{6};
\frac{-1}{2},\frac{5}{6},\frac{-1}{6}\right)
\left(\frac{-1}{2},\frac{1}{6},\frac{-1}{3};0^5\right)'$ & ${\bf
1}_{1/3}(T_2^-)$
%
& $\overline{\eta}_{7}$
\\
$\left(\frac{1}{6},\frac{1}{6},\frac{1}{6};\frac{-1}{6},\frac{-1}{6};
\frac{1}{2},\frac{-1}{6},\frac{-1}{6}\right)
\left(\frac{1}{2},\frac{1}{6},\frac{2}{3};0^5\right)'$ & ${\bf
1}_{1/3}(T_2^-)$
%
& $\overline{\eta}_{8}$
\\
$\left(\frac{1}{6},\frac{1}{6},\frac{1}{6};\frac{-1}{6},\frac{-1}{6};
\frac{1}{2},\frac{-1}{6},\frac{-1}{6}\right)
\left(\frac{-1}{2},\frac{1}{6},\frac{-1}{3};0^5\right)'$ &$2\cdot
{\bf 1}_{1/3}(T_2^-)$
%
&
$\overline{\eta}_{9}, \overline{\eta}_{10}$
\\
$\left(\frac{-1}{3},\frac{-1}{3},\frac{-1}{3};\frac{1}{3},\frac{1}{3};
\frac{-2}{3},\frac{-1}{3},0\right)
\left(0,\frac{-1}{3},\frac{-1}{3};0^5\right)'$ & $2\cdot {\bf
1}_{-2/3}(T_4^+)$
%
&  $\overline{\xi}_{4}$
\\
$\left(\frac{-1}{3},\frac{-1}{3},\frac{-1}{3};\frac{1}{3},\frac{1}{3};
\frac{1}{3},\frac{2}{3},0\right)
\left(0,\frac{-1}{3},\frac{-1}{3};0^5\right)'$ &  $2\cdot {\bf
1}_{-2/3}(T_4^+)$
%
& $\overline{\xi}_{5}$
\\
$\left(\frac{1}{6},\frac{1}{6},\frac{1}{6};\frac{-1}{6},\frac{-1}{6};
\frac{-1}{6},\frac{-5}{6},\frac{-1}{2}\right)
\left(0,\frac{-1}{3},\frac{-1}{3};0^5\right)'$ & $3\cdot {\bf
1}_{1/3}(T_4^+)$
%
& $\overline{\eta}_{11}$
\\
$\left(\frac{1}{6},\frac{1}{6},\frac{1}{6};\frac{-1}{6},\frac{-1}{6};
\frac{5}{6},\frac{1}{6},\frac{-1}{2}\right)
\left(0,\frac{-1}{3},\frac{-1}{3};0^5\right)'$ & $2\cdot {\bf
1}_{1/3}(T_4^+)$
%
& $\overline{\eta}_{12}$
\\
$\left(\frac{1}{6},\frac{1}{6},\frac{1}{6};\frac{-1}{6},\frac{-1}{6};
\frac{-1}{6},\frac{1}{6},\frac{1}{2}\right)
\left(0,\frac{2}{3},\frac{2}{3};0^5\right)'$ &$2\cdot {\bf
1}_{1/3}(T_4^+)$
%
& $\overline{\eta}_{13}$
\\
$\left(\frac{1}{6},\frac{1}{6},\frac{1}{6};\frac{-1}{6},\frac{-1}{6};
\frac{-1}{6},\frac{1}{6},\frac{1}{2}\right)
\left(0,\frac{-1}{3},\frac{-1}{3};0^5\right)'$ & $6\cdot {\bf
1}_{1/3}(T_4^+)$
%
&
$\overline{\eta}_{14}, \overline{\eta}_{15}, \overline{\eta}_{16}$
\\
$\left(\frac{1}{3},\frac{1}{3},\frac{1}{3};\frac{-1}{3},\frac{-1}{3};
0,\frac{2}{3},\frac{-1}{3}\right)
\left(0,\frac{1}{3},\frac{1}{3};0^5\right)'$ & $2\cdot {\bf
1}_{2/3}(T_4^-)$
%
& $\xi_{6}$
\\
$\left(\frac{1}{3},\frac{1}{3},\frac{1}{3};\frac{-1}{3},\frac{-1}{3};
0,\frac{-1}{3},\frac{2}{3}\right)
\left(0,\frac{1}{3},\frac{1}{3};0^5\right)'$ & $2\cdot {\bf
1}_{2/3}(T_4^-)$
%
& $\xi_{7}$
\\
$\left(\frac{-1}{6},\frac{-1}{6},\frac{-1}{6};\frac{1}{6},\frac{1}{6};
\frac{-1}{2},\frac{-5}{6},\frac{1}{6}\right)
\left(0,\frac{1}{3},\frac{1}{3};0^5\right)'$ & $3\cdot {\bf
1}_{-1/3}(T_4^-)$
%
& $\eta_{17}$
\\
$\left(\frac{-1}{6},\frac{-1}{6},\frac{-1}{6};\frac{1}{6},\frac{1}{6};
\frac{-1}{2},\frac{1}{6},\frac{-5}{6}\right)
\left(0,\frac{1}{3},\frac{1}{3};0^5\right)'$ & $3\cdot {\bf
1}_{-1/3}(T_4^-)$
%
&  $\eta_{18}$
\\
$\left(\frac{-1}{6},\frac{-1}{6},\frac{-1}{6};\frac{1}{6},\frac{1}{6};
\frac{1}{2},\frac{1}{6},\frac{1}{6}\right)
\left(0,\frac{-2}{3},\frac{-2}{3};0^5\right)'$ & $2\cdot {\bf
1}_{-1/3}(T_4^-)$
%
& $\eta_{19}$
\\
$\left(\frac{-1}{6},\frac{-1}{6},\frac{-1}{6};\frac{1}{6},\frac{1}{6};
\frac{1}{2},\frac{1}{6},\frac{1}{6}\right)
\left(0,\frac{1}{3},\frac{1}{3};0^5\right)'$ & $6\cdot {\bf
1}_{-1/3}(T_4^-)$
%
&
$\eta_{20},\eta_{21},\eta_{22}$
\\

 $\left(\frac{-1}{12},\frac{-1}{12},\frac{-1}{12};
\frac{7}{12},\frac{7}{12};
\frac{1}{12},\frac{-7}{12},\frac{-3}{12}\right)
\left(\frac{3}{12},\frac{5}{12},\frac{-4}{12};0^5\right)'$ & ${\bf
1}_{-2/3}(T_1^+)$
%
& $\overline{\xi}_{8}$
\\
$\left(\frac{-1}{12},\frac{-1}{12},\frac{-1}{12};
\frac{-5}{12},\frac{-5}{12};
\frac{1}{12},\frac{5}{12},\frac{9}{12}\right)
\left(\frac{3}{12},\frac{5}{12},\frac{-4}{12};0^5\right)'$ & ${\bf
1}_{1/3}(T_1^+)$
%
&$\overline{\eta}_{23}$
\\
$\left(\frac{5}{12},\frac{5}{12},\frac{5}{12};
\frac{1}{12},\frac{1}{12};
\frac{7}{12},\frac{-1}{12},\frac{3}{12}\right)
\left(\frac{3}{12},\frac{5}{12},\frac{-4}{12};0^5\right)'$ &${\bf
1}_{1/3}(T_1^+)$
%
& $\overline{\eta}_{24}$
\\
$\left(\frac{-1}{12},\frac{-1}{12},\frac{-1}{12};
\frac{-5}{12},\frac{-5}{12};
\frac{1}{12},\frac{-7}{12},\frac{-3}{12}\right)
\left(\frac{3}{12},\frac{5}{12},\frac{-4}{12};0^5\right)'$ & ${\bf
1}_{1/3}(T_1^+)$
%
& $\overline{\eta}_{25}$
\\
$\left(\frac{1}{12},\frac{1}{12},\frac{1}{12};
\frac{5}{12},\frac{5}{12};
\frac{-9}{12},\frac{-1}{12},\frac{-1}{12}\right)
\left(\frac{3}{12},\frac{1}{12},\frac{-8}{12};0^5\right)'$ & ${\bf
1}_{-1/3}(T_1^-)$
%
& $\eta_{26}$
\\
$\left(\frac{-1}{12},\frac{-1}{12},\frac{-1}{12};
\frac{7}{12},\frac{7}{12};
\frac{1}{12},\frac{5}{12},\frac{-3}{12}\right)
\left(\frac{-3}{12},\frac{-1}{12},\frac{8}{12};0^5\right)'$ & ${\bf
1}_{-2/3}(T_7^+)$
%
& $\overline{\xi}_{9}$
\\
$\left(\frac{5}{12},\frac{5}{12},\frac{5}{12};
\frac{1}{12},\frac{1}{12};
\frac{-5}{12},\frac{-1}{12},\frac{3}{12}\right)
\left(\frac{9}{12},\frac{-1}{12},\frac{-4}{12};0^5\right)'$ & ${\bf
1}_{1/3}(T_7^+)$
%
& $\overline{\eta}_{27}$
\\
$\left(\frac{5}{12},\frac{5}{12},\frac{5}{12};
\frac{1}{12},\frac{1}{12};
\frac{-5}{12},\frac{-1}{12},\frac{3}{12}\right)
\left(\frac{-3}{12},\frac{-1}{12},\frac{8}{12};0^5\right)'$ &${\bf
1}_{1/3}(T_7^+)$
%
&$\overline{\eta}_{28}$
\\
$\left(\frac{-1}{12},\frac{-1}{12},\frac{-1}{12};
\frac{-5}{12},\frac{-5}{12};
\frac{1}{12},\frac{5}{12},\frac{-3}{12}\right)
\left(\frac{9}{12},\frac{-1}{12},\frac{-4}{12};0^5\right)'$ &${\bf
1}_{1/3}(T_7^+)$
%
& $\overline{\eta}_{29}$
\\
$\left(\frac{-1}{12},\frac{-1}{12},\frac{-1}{12};
\frac{-5}{12},\frac{-5}{12};
\frac{1}{12},\frac{5}{12},\frac{-3}{12}\right)
\left(\frac{-3}{12},\frac{-1}{12},\frac{8}{12};0^5\right)'$ &${\bf
1}_{1/3}(T_7^+)$
%
& $\overline{\eta}_{30}$
\\
$\left(\frac{1}{12},\frac{1}{12},\frac{1}{12};
\frac{-7}{12},\frac{-7}{12};
\frac{3}{12},\frac{-1}{12},\frac{-1}{12}\right)
\left(\frac{-3}{12},\frac{-5}{12},\frac{4}{12};0^5\right)'$ &${\bf
1}_{2/3}(T_7^-)$
%
& ${\xi}_{10}$
\\
$\left(\frac{1}{12},\frac{1}{12},\frac{1}{12};
\frac{5}{12},\frac{5}{12};
\frac{3}{12},\frac{-1}{12},\frac{-1}{12}\right)
\left(\frac{9}{12},\frac{7}{12},\frac{4}{12};0^5\right)'$ &${\bf
1}_{-1/3}(T_7^-)$
%
& ${\eta}_{31}$
\\
$\left(\frac{1}{12},\frac{1}{12},\frac{1}{12};
\frac{5}{12},\frac{5}{12};
\frac{3}{12},\frac{-1}{12},\frac{-1}{12}\right)
\left(\frac{-3}{12},\frac{7}{12},\frac{-8}{12};0^5\right)'$ &${\bf
1}_{-1/3}(T_7^-)$
%
& ${\eta}_{32}$
\\
$\left(\frac{-5}{12},\frac{-5}{12},\frac{-5}{12};
\frac{-1}{12},\frac{-1}{12};
\frac{-3}{12},\frac{5}{12},\frac{5}{12}\right)
\left(\frac{-3}{12},\frac{-5}{12},\frac{4}{12};0^5\right)'$ &${\bf
1}_{-1/3}(T_7^-)$
%
& ${\eta}_{33}$
\\
$\left(\frac{1}{12},\frac{1}{12},\frac{1}{12};
\frac{5}{12},\frac{5}{12};
\frac{3}{12},\frac{-1}{12},\frac{-1}{12}\right)
\left(\frac{-3}{12},\frac{-5}{12},\frac{4}{12};0^5\right)'$ &$2\cdot
{\bf 1}_{-1/3}(T_7^-)$
%
& ${\eta}_{34},
{\eta}_{35}$
\\[0.2em]
\hline
\end{tabular}
\end{center}
 \caption{Singlet exotics. }\label{tb:Sexotics}
 }
\end{table}
In these tables, $\xi,\overline{\xi}$ are \Qem$=\pm\frac23$
singlets and $\eta,\overline{\eta}$ are \Qem$=\mp\frac13$
singlets. Singlet exotics of Table \ref{tb:Sexotics} are
vectorlike.

We find that fields with non-vanishing U(1)$_6$ quantum numbers are
only exotics. This means that U(1)$_6$ cannot be broken by VEVs of
neutral singlets since neutral singlets cannot be exotics. As
mentioned before, however, the exactly massless U(1)$_6$ gauge boson
(``exphoton'') is still phenomenologically acceptable, since all
observable matter fields are neutral under U(1)$_6$.

In Table \ref{tb:singexmass}, we present some mass terms of singlet
exotics. In this mass table, we tried to combine vectorlike pairs,
not listing all off-diagonal terms as before. It would be unwieldy
to list all the off-diagonal terms for several tens of singlets.
\begin{table}[h]
\begin{center}
\begin{tabular}{|c|c|}
\hline Pairs & ~Masses ($\times$proper CS)
\\ \hline &\\[-1.2em]
$1\times\{\xi_1(T_2^+) ,~ \overline{\xi}_9(T_7^+)\}$ &
$S_{7}S_{9}$\fbox{$S_{23}$}
\\
$1\times\{\xi_2(T_2^+) ,~ \overline{\xi}_3(T_2^-)\}$ & $S_1S_{10}$
\\
$1\times\{\xi_{10}(T_7^-) ,~ \overline{\xi}_8(T_1^+)\}$ & $S_8$,
$S_2S_4$
\\
$2\times\{\xi_6(T_4^-) ,~ \overline{\xi}_4(T_4^+)\}$ & $S_9$,
$S_3S_4$
\\
$2\times\{\xi_7(T_4^-) ,~ \overline{\xi}_5(T_4^+)\}$ & $S_7$,
$S_1S_5$
\\
\hline $1\times\{\eta_1(T_2^+) ,~ \overline{\eta}_6(T_2^-)\}$ &
$S_4S_{10}$
\\
$1\times\{\eta_2(T_2^+) ,~ \overline{\eta}_7(T_2^-)\}$ &
$S_1S_4S_9$
%
%
\\
$1\times\{\eta_3(T_2^+) ,~ \overline{\eta}_8(T_2^-)\}$ &
$S_5S_{11}$
\\
$\{\eta_{4,5}(T_2^+) ,~ \overline{\eta}_{9,10}(T_2^-)\}$ &
$S_5S_{11}$
\\
$2\times\{\eta_{17}(T_4^-) ,~ \overline{\eta}_{12}(T_4^+)\}$ &
$S_8$, $S_2S_4$
\\
$1\times\{\eta_{17}(T_4^-) ,~ \overline{\eta}_{27}(T_7^+)\}$ &
$S_6S_{29}$
\\
$3\times\{\eta_{18}(T_4^-) ,~ \overline{\eta}_{11}(T_4^+)\}$ &
$(S_4S_{12})^2$
\\
$2\times\{\eta_{19}(T_4^-) ,~ \overline{\eta}_{13}(T_4^+)\}$ &
$S_7$, $S_1S_5$
\\
$\{\eta_{20,21,22}(T_4^-) ,~ \overline{\eta}_{14,15,16}(T_4^+)\}$
& $S_7$, $S_1S_5$
\\
$1\times\{\eta_{26}(T_1^-) ,~ \overline{\eta}_{30}(T_7^+)\}$ &
$S_9$, $S_3S_3$
\\
$1\times\{\eta_{31}(T_7^-) ,~ \overline{\eta}_{29}(T_7^+)\}$ &
$S_1S_7S_{12}$\fbox{$S_{15}$} $S_{29}$
\\ [0.2em]
$1\times\{\eta_{32}(T_7^-) ,~ \overline{\eta}_{28}(T_7^+)\}$ &
$S_{13}$\fbox{$S_{23}$} $S_{29}$
\\
$1\times\{\eta_{33}(T_7^-) ,~ \overline{\eta}_{24}(T_1^+)\}$ &
$S_7$, $S_1S_5$
\\
$1\times\{\eta_{34}(T_7^-) ,~ \overline{\eta}_{23}(T_1^+)\}$ &
$S_7$, $S_1S_5$
\\
$1\times\{\eta_{35}(T_7^-) ,~ \overline{\eta}_{25}(T_1^+)\}$ &
$S_8$, $S_2S_4$
\\
\hline
\end{tabular}
\end{center}
\caption{Mass terms for singlet exotics. CS are products of
singlet fields given in Eq. (2.19). Proper CS are assumed to be
multiplied such that the $H$-momentum becomes $(-1,1,1)$ mod
$(12,3,12)$. We set $\langle {\rm CS}\rangle=1$.}
 \label{tb:singexmass}
\end{table}
We note that in the above vacuum (\ref{VEVsinglets}), all exotic
singlets obtain masses, as can be seen from the pairings listed in
Table \ref{tb:singexmass}. But here it seems that $\Gamma$ odd
fields $S_{15}$ and $S_{23}$ are involved.

\section{$D$ and $F$ flat directions}\label{sec:DFflat}

\subsection{Anomalous U(1) and $D$ flat directions}

There are eight U(1) symmetries in this model. If there is an
anomalous U(1), some of the gauge symmetries are broken via the
Fayet-Iliopoulos $D$-term. Indeed, our model has an anomalous
U(1)$_A$ whose charge is given in terms of  the original eight U(1)
 charges as
\begin{eqnarray}
Q_A = 24Y-30(B-L) +Q_1 + Q_2 + Q_3 + Q_4 - Q_5 .
\end{eqnarray}
The Fayet-Iliopoulos $D$-term is
\begin{equation}
D^{A}= \frac{2g}{192\pi^2}{\rm Tr}Q_A+\sum_i
Q_A(i)\phi^*(i)\phi(i) .\label{FYDterm}
\end{equation}
As shown in Appendix \ref{App:B}, Tr$Q_A$ is negative, $-50$.  For
supersymmetry, the chosen vacuum must satisfy $\langle
D^A\rangle=0$. Thus the summation
%
$\sum_i Q_A(i)\phi^*(i)\phi(i)$ for the nonzero VEVs given in
(\ref{VEVsinglets}) should be positive. The VEVs in $D^A$ term
potential can break a U(1) at the SUSY minimum. To see how the
remaining six U(1)s behave, in Table \ref{U(1)VEVs} we list the
U(1) charges of those singlets with non-vanishing VEVs. The
$D$-flatness conditions for the remaining anomaly free U(1)$_g$
are
\begin{equation}
\langle D^{(g)}\rangle= \left\langle \sum_i Q_g(i)\phi^*(i)\phi(i)
\right\rangle=0, \quad g=Y,~a,~b,~\cdots,~e,~6.
\end{equation}


\begin{table}
\begin{center}
\begin{tabular}{|c||c|c|c|c|c|c|c|c|c|}
\hline  Label   & ${\cal P}(f_0)$ & $ Q_Y$ & $Q_{A}$ & $ Q_a$ & $
Q_b$ & $ Q_c$ & $ Q_d$ & $ Q_e$ & $ Q_6$
\\ \hline
\hline   $S_0(U_2)$  & 1 & 0 & $0$ & $2$ & $6$ & $0$ & $0$ & $0$ & 0
\\
\hline $S_{1}(T_2^0)$  & 1 & 0 &  $\frac{-4}{3}$ & $0$ &
$\frac{-4}{3}$ & $\frac{-19}{3}$ & $\frac{-19}{3}$ & $\frac{-4}{3}$
& 0

\\
$S_{2}(T_2^0)$   & 1 & 0 & $\frac{-4}{3}$ & $-2$ & $\frac{-10}{3}$ &
$\frac{11}{3}$ & $\frac{11}{3}$ & $\frac{-4}{3}$ & 0
\\
$S_{3}(T_2^0)$   & 1 & 0 &  $\frac{-4}{3}$ & $2$ & $\frac{-10}{3}$ &
$\frac{11}{3}$ & $\frac{11}{3}$ & $\frac{-4}{3}$ & 0
\\
$S_{4}(T_2^0)$  & 1+1 & 0 & $\frac{8}{3}$ & $0$ & $\frac{8}{3}$ &
$\frac{-7}{3}$ & $\frac{-7}{3}$ & $\frac{8}{3}$ & 0
\\
$S_{5}(T_2^0)$  & 1+1 & 0 & $\frac{-4}{3}$ & $0$ & $\frac{8}{3}$ &
$\frac{11}{3}$ & $\frac{11}{3}$ & $\frac{-4}{3}$ & 0
\\
\hline $S_{6}(T_4^0)$   & 2 & 0 & $\frac{4}{3}$ & $0$ &
$\frac{16}{3}$ & $\frac{4}{3}$ & $\frac{4}{3}$ & $\frac{4}{3}$ & 0
\\
$S_{7}(T_4^0)$  & 2+3+2 & 0 & $\frac{-8}{3}$ & $0$ & $\frac{4}{3}$ &
$\frac{-8}{3}$ & $\frac{-8}{3}$ & $\frac{-8}{3}$ & 0
\\
$S_{8}(T_4^0)$   & 2+2+2 & 0 & $\frac{4}{3}$ & $-2$ & $\frac{-2}{3}$
& $\frac{4}{3}$ & $\frac{4}{3}$ & $\frac{4}{3}$ & 0
\\
$S_{9}(T_4^0)$  & 2+2+2 & 0 & $\frac{4}{3}$ & $2$ & $\frac{-2}{3}$ &
$\frac{4}{3}$ & $\frac{4}{3}$ & $\frac{4}{3}$ & 0
\\
\hline  $S_{10}(T_6)$  & 2 & 0 & $0$ & $2$ & $2$ & $5$ & $5$ & $0$ &
0
\\
$S_{11}(T_6)$   & 2 & 0 & $0$ & $-2$ & $-2$ & $-5$ & $-5$ & $0$ & 0
\\
$S_{12}(T_6)$   & 2 & 0 & $0$ & $0$ & $-4$ & $5$ & $5$ & $0$ & 0
\\
$S_{13}(T_6)$   & 2 & 0 & $0$ & $0$ & $4$ & $-5$ & $-5$ & $0$ & 0
\\
\hline
 $S_{15}(T_1^0)$  & 1 &  0 &  $\frac{-85}{6}$ & $0$ &
$\frac{4}{3}$ & $\frac{19}{3}$ & $\frac{-5}{3}$ & $\frac{5}{4}$ &
0
\\
$S_{23}(T_7^0)$  & 1 & 0 & $\frac{101}{6}$ & $0$ & $\frac{-8}{3}$
& $\frac{-11}{3}$ & $\frac{13}{3}$ & $\frac{17}{12}$ & 0
\\
$S_{29}(T_9)$   & 2 & 0 &  $\frac{31}{2}$ & $0$ & $0$ & $6$ & $-2$
& $\frac{1}{12}$ & 0
\\
\hline
\end{tabular}
\end{center}
\caption{U(1) charges of scalars developing nonzero
VEVs.}\label{U(1)VEVs}
\end{table}

One could find the solution to $D^A=D^{(g)}=0$
($g=Y,~a,~b,~\cdots,~e,~6$):
\begin{eqnarray}
|S_0|^2&=&|S_{4}|^2 -2|S_{5}|^2 -|S_{6}|^2 -7|S_{7}|^2 +3|S_{8}|^2
+3|S_{9}|^2
\\ \label{sol0}\nonumber
&&-|S_{10}|^2 +|S_{11}|^2 +|S_{12}|^2 -|S_{13}|^2
+|S_{23}|^2-\frac{7X^2}{1480} ,
\\
|S_1|^2&=&|S_{4}|^2 +|S_{6}|^2 -7|S_{7}|^2 +3|S_{8}|^2 +3|S_{9}|^2
\\ \label{sol1}\nonumber
&&+|S_{10}|^2 -|S_{11}|^2 +|S_{12}|^2 -|S_{13}|^2 +|S_{23}|^2
-\frac{17X^2}{1480} ,
\\
|S_2|^2&=&2|S_{4}|^2 -2|S_{5}|^2 -7|S_{7}|^2 +6|S_{9}|^2
+|S_{23}|^2 -\frac{2X^2}{1480} ,
\label{ sol2} \\
|S_3|^2&=&|S_{4}|^2 +|S_{6}|^2 +3|S_{8}|^2 -3|S_{9}|^2 -|S_{10}|^2
+|S_{11}|^2 -|S_{12}|^2 +|S_{13}|^2 -\frac{9X^2}{1480} ,
\label{sol3} \\
|S_{15}|^2&=&|S_{23}|^2-\frac{6X^2}{185}
\label{sol15}\\
|S_{29}|^2&=&\frac{3X^2}{185} \label{sol29},
\end{eqnarray}
where $X^2\equiv \frac{-2g}{192\pi^2}{\rm Tr}Q_A$. Eq.
(\ref{sol29}) dictates $S_{29}\sim {\cal O}(M_{\rm string}/100)$.
The following hierarchical assumption for the VEVs could be
consistent with Eq. (\ref{sol0}), (\ref{sol1}), and (\ref{sol3}):
\begin{eqnarray}
\frac12|S_{3}|^2\approx|S_6|^2 \approx |S_{11}|^2 ~\gtrsim
~{\rm others} .
\end{eqnarray}

As we mentioned before U(1)$_6$ remains unbroken since there is no
neutral singlet carrying a nonzero $Q_6$. Thus, in addition to
photon there exists another strictly massless U(1)$_6$ gauge boson
({\it exphoton}).   It couples only to superheavy exotic matter.

\subsection{$F$ flat directions}

The neutral singlets in Table \ref{tb:Singneutral} classified to
the five categories as shown in Table \ref{4class}.
\begin{table}
\begin{center}
\begin{tabular}{|c||c|c|c|}
\hline Classes  & $\Gamma'$ & VEV & Neutral singlets
\\  \hline
I & $0$ & non-zero& $S_0,~S_1,~S_2,~\cdots ,~S_{13}
,~S_{15},~S_{23}$
\\
II & $+1$ & zero &
$S_{14},~S_{19},~S_{21},~S_{25},~S_{27},~S_{28}$
\\
III & $-1$ & (non-)zero & $S_{17},~S_{22},~S_{24},~S_{26}$
\\
IV & $0$ & (non-)zero & $S_{16},~S_{18},~S_{20}$
\\
V & $-1$ & non-zero & $S_{29}$
\\
\hline
\end{tabular}
\end{center}
\caption{Five classes of the neutral singlets.}\label{4class}
\end{table}
The singlets included in Class I, which do not carry
U(1)$_{\Gamma'}$ charges defined in Eq. (\ref{Gamma}), are assumed
to develop VEVs. The singlets in Class V are also assumed to get
VEVs, but they carry the U(1)$_{\Gamma}$ charges of $-1$. On the
other hand, the singlet states in Classes II, III, and IV which
carry $\Gamma'=\pm 1$ or $0$, do not obtain VEVs.

We note that the $R$-parity violating operators, $u^cd^cd^c$,
$QLd^c$, $LLe^c$ carry $\Gamma'=-1$. Thus, if VEVs by singlets
carrying positive $\Gamma'$ charges are absent, as in our case,
the trilinear $R$-parity violating terms could not be induced in
the superpotential. Hence, if necessary, the singlets in III and
IV, which all have the zero or negative $\Gamma'$ charges, can be
allowed to get VEVs. In this paper, however, for simplicity we
consider only a vacuum where all singlets in the classes III and
IV do not obtain VEVs.

There exist superpotential terms constructed purely with the
neutral singlet fields in the class I:
\begin{eqnarray}
&&W = S_1S_6S_{12}+S_3S_6S_{11}+S_1S_8S_{10}+S_3S_8S_{13}
+S_2S_9S_{13}
\nonumber\\
&&~~+S_4S_7S_{12}+S_5S_9S_{11}+S_7S_8S_9+S_{10}S_{11}+S_{12}S_{13}
+\cdots ,
\end{eqnarray}
where proper CS are assumed to be multiplied. As seen in Eq.
(\ref{CSHmom}), CS are constructed also with the singlets in Class
I. In the ${\bf Z}_{12}$ orbifold compactification, if a
superpotential term $w$ satisfies all the selections rules, then
$w^{12n+1}$ ($n=1,2,3, \cdots$) also does. By including the higher
dimensional terms $w^{13}$, $w^{25}$, $w^{37}, \cdots$, one can
always find a vacuum where the singlets of interest develop VEVs of
string scale, preserving the $F$ flatness conditions \cite{Buch}.
Moreover, one can always find a re-scaling transformation for the
VEVs, leaving intact the $F$ flatness conditions. Using this
transformation, one can be consistent also the $D$ flatness
conditions can be consistent \cite{Buch,Nilles06}. With this
justification we assume that all the neutral singlets of the class I
achieve VEVs of order $M_{\rm string}$ on a vacuum. As argued
earlier, the selection rule Eq. (\ref{Hconsv}) reduces to Eq.
(\ref{simpleHmom}) on such a vacuum.

Yukawa couplings containing two or more singlets with zero VEVs
are trivial in satisfying the $F$-flatness conditions. Thus, the
couplings, in which more than two singlets out of II, III or IV
are involved, do not provide non-trivial constraints for
$F$-flatness. However, in the presence of a coupling including
only one singlet with vanishing VEV, $F$-flatness may not be
present unless there are more than two such terms.

In the superpotential, the singlets should couple to other fields
such that Yukawa couplings are neutral under U(1)$_{\Gamma'}$ and
also the other U(1) gauge symmetries: $\Gamma'$ charges of
singlets in the class III should be compensated by being coupled
with those of singlets in II. Since all the singlets in II and III
do not get VEVs, the couplings between II and III do not provide
non-trivial constraints for $F$-flatness. On the other hand, we
should be careful for the couplings between singlets in I and IV,
and in II and V, because in these cases couplings only one singlet
with a vanishing VEV are possible.  In this model, indeed, one can
find two or more allowed superpotential terms for each singlet in
II.
Therefore, the $F$-flatness conditions, $\partial W/\partial
S_{14}=\partial W/\partial S_{16}=\partial W/\partial
S_{18}=\cdots =\partial W/\partial S_{28}=0$ can be satisfied.
$D$-flatness conditions can be satisfied by re-scaling of VEVs.
However, in order to get $\langle S_{14}\rangle=\langle
S_{16}\rangle =\langle S_{18}\rangle=\cdots =\langle
S_{28}\rangle=0$ as a $F$-flatness solution and also a $\mu$
solution, many $F$-flatness conditions should turn out not be
independent ones.

\section{Vacuum with effective $R$ parity}\label{sec:Rparity}

For the $R$-parity to be exact, it must be a subgroup of a U(1)
gauge group, i.e. it must be  a discrete gauge symmetry
\cite{discreteg}, otherwise large gravitational corrections such
as through wormhole processes may violate it. Here, we can include
the anomalous U(1) gauge symmetry in string compactification
\cite{Dine:1987xk}, since the matter anomaly is cancelled by the
Green-Schwarz mechanism \cite{Green:1984sg}. Taking out the SM
nonabelian gauge groups from the E$_8$ sector leaves five U(1)s
among which U(1)$_Y$ cannot be used for the $R$-parity. Thus, for
the $R$-parity,  we are left with four possibilities,
\begin{equation}
\begin{array}{ccccccccccccc}
(B-L)&=& (&\frac23&\frac23&\frac23&0&0&0&0&0&)&(0^8)'\\
X&=& (&-2&-2&-2&-2&-2&0&0&0&)&(0^8)'\\
Q_1&=& (&0&0&0&0&0&+2&0&0&)&(0^8)'\\
Q_2&= &(&0&0&0&0&0&0&+2&0&)&(0^8)'
\end{array}\label{Rparfs}
\end{equation}
For example, another U(1) charge $(-2,-2,-2,-2,-2,-2,-2,0)(0^8)' $
is the linear combination $X-Q_1-Q_2$. For an $R$-parity, we can
use any odd number of U(1)s given in Eq. (\ref{Rparfs}). The
reason is the following. It is customarily assumed that the SO(10)
subgroup of E$_8$ allows the spinor representation of SO(10). If
it arises in the untwisted sector, it must be of the form
\begin{equation}
(\underline{[+++++]};\underline{[+++]})\label{rep:spin}
\end{equation}
where $\pm$ are $\pm\frac12$, and the underline means all possible
permutations and [ ] means  even numbers of  sign flips among
entries inside the bracket. For the representation (\ref{rep:spin}),
the U(1) charges of (\ref{Rparfs}) are odd. On the other hand, the
Higgs doublets in SO(10) have the form
\begin{equation}
(0~0~0~\underline{\pm1~0};\pm1~0)\label{rep:vec}
\end{equation}
which give even numbers of the U(1) charges of (\ref{Rparfs}). We
can define a good $R$-parity if all the scalar fields developing
VEVs carry even numbers of a U(1) charge, say $\Gamma$, of
(\ref{Rparfs}). Here, a conflict arises if the phenomenologically
needed VEVs require for some $\Gamma$ odd fields to develop VEVs.
Then, in general an exact parity cannot be defined.

Let us note possible superpotential terms in the MSSM, generating
$\Delta B\ne 0$ operators,
\begin{align}
&d=4:\quad u^cd^cd^c, \label{D3uud}\\
&d=5:\quad QQQL,\quad u^cu^cd^ce^c\label{D3qqql}
\end{align}
where $Q$ and $L$ are quark and lepton doublets, respectively. The
dimension-4 operator of Eq. (\ref{D3uud}) alone  does not lead to
proton decay, but that term together with the $\Delta L\ne 0$
superpotential $QLd^c$ leads to a very fast proton decay and the
product of their couplings must satisfy a very stringent
constraint, $<10^{-26}$. The $d=5$ operators in (\ref{D3qqql}) are
not that much dangerous, but still the couplings must satisfy
constraints, $<10^{-7}$ \cite{SakYan,ibross}. Thus, our prime
objective of introducing the $R$-parity is to forbid $u^cd^cd^c$
up to a sufficiently high level.

A $\Z_2$ subgroup of a U(1) gauge symmetry is welcome for a
definition of $R$-parity. The continuous global U(1) symmetry,
being broken by superpotential terms, is not good for an
$R$-parity. For this, we note that the $\Z_2$ subgroup of the
U(1)$_X$ gauge group distinguishes the spinor or the vector origin
of our spectrum where
\begin{equation}
X=(-2,-2,-2,-2,-2,~0,~0,~0)(0^8)'.
\end{equation}
For distinguishing two kinds of parity quantum numbers in our model,
actually we have a better U(1) gauge symmetry, U(1)$_\Gamma$, whose
generator is
\begin{equation}
\Gamma=\textstyle X -(Q_2+Q_3)+\frac14(Q_4+Q_5)
+6(B-L)\label{U1Gamma}
\end{equation}
where
\begin{equation}
\begin{array}{l}
Q_2=(0^5;0,2,0)(0^8)',\quad  Q_3=(0^5;0,0,2)(0^8)'\\
 Q_4=(0^8)(2,0,0^6)',\quad Q_5=(0^8)(0,2,0^6)'\\
 B-L=(\frac23,\frac23,\frac23,0^5)(0^8)'
 \end{array}.
\end{equation}
$Q_4$ and $Q_5$ in (\ref{U1Gamma}) affect only the hidden E$_8'$.
In Eq. (\ref{U1Gamma}), there is an odd number of operators of Eq.
(\ref{Rparfs}), and hence $\Gamma$ is good for defining a parity.
The $\Gamma$ quantum numbers of standard charge particles are
listed in Tables \ref{tb:standardch} and \ref{tb:Singneutral}. Let
us define the $R$-parity by giving VEVs to some $\Gamma=\pm 2,0$
neutral singlets,
\begin{equation}
U(1)_\Gamma\longrightarrow Z_2\equiv P.\label{defRP}
\end{equation}
The parity defined in this way is multiplicative. Then, the even
integer fields carry $P=+1$ and the odd integer fields carry $P=-1$.
The $P$ allowed couplings must have the {\it total} $P=+1$. A more
restrictive condition is the U(1)$_\Gamma$ gauge invariance of
couplings: $\sum_i \Gamma(z_i)=0$, which must be satisfied for the
coupling to be present in the original theory.

Inspecting the $\Gamma$ quantum numbers in the tables, we find
that the following fields are possessing $\lq$strange' $\Gamma$s
in defining the $R$-parity:
\begin{align}
&\overline{D}(T_4^0),\quad {D}(T_4^0),\quad
\overline{D}(T_6),\quad {D}(T_6),\label{Dstrange}\\
&~~~S_{15},\quad S_{16} ,\quad S_{18},\quad S_{20},\quad
S_{23}\label{Sstrange}
\end{align}
which are boxed in Tables and \ref{tb:standardch} and
\ref{tb:Singneutral}. Fields in (\ref{Dstrange}) carry the
familiar charge \Qem$=-\frac13$ down type charge but carry even
$\Gamma$s, \Qem$=0$ neutral singlets in (\ref{Sstrange}) are the
familiar neutral Higgs fields but they carry odd $\Gamma$s. To
have an exact $R$-parity, neutral singlets $S_{15}$, $S_{16}$,
$S_{18}$, $S_{20}$, and $S_{23}$ (the boxed ones) in Table
\ref{tb:Singneutral} should not develop VEVs. But then, the
leftover pair in Table \ref{tb:DDbarmass} cannot obtain mass since
$\overline{D}(T_4^0)$ carries $P=2$ and $D(T_9)$ carries $P=-1$.
To give them mass, some of $S_{15}$, $S_{16}$, $S_{18}$, $S_{20}$,
and $S_{23}$ should develop VEV(s). These VEVs violate the
$R$-parity, i.e. $P$. So, in our model $R$-parity violation is
inevitable to give large masses to exotics.


\subsection{$R$ parity violation}\label{subsec:Rviol}
As mentioned above, the dimension-5 operators of the form $QQQL$
and $u^cu^cd^ce^c$, allowed by $R$-parity, are known to be safe
for the proton lifetime constraint in string compactification
models \cite{KimRapp}. To constrain the $R$-parity violation from
the $\Delta B\ne 0$ processes, therefore, we focus on dimension-4
superpotential terms of the form $u^cd^cd^c$ attached with some of
$S_{15}$, $S_{16}$, $S_{18}$, $S_{20}$, and $S_{23}$. If there
does not exist any such term, the $R$-parity violation is safe
from the proton lifetime bound. The mixing of
$\overline{D}(T_4^0)$ with $d^c$ is $O( 10^{-16})$ for
$m_{\overline{D}(T_4^0)}= O(10^{16})$ GeV, and hence we will not
consider the $R$-parity preserving coupling,
$u^cd^c\overline{D}(T_4^0)$.

To study the non-renormalizable couplings, we need products of
singlets having non-vanishing VEVs, shown in Eq.
(\ref{VEVsinglets}). Among these, non-vanishing $\Gamma$s are
carried by $S_0(\Gamma=2), S_{15}(T_1^0,\Gamma=-1)$, and $
S_{23}(T_7^0,\Gamma=-1), S_{29}(T_9,\Gamma=-2)$. Since $u^cd^cd^c$
carries $\Gamma=-3$, we need singlet products having $\Gamma=+3$. So
we must satisfy two conditions: inclusion of $S_0$ and inclusion of
an odd number of $ S_{15}$ and $ S_{23}$. Of course, the
$H$-momentum rules and the gauge invariance conditions must be
satisfied. Let us consider the following example of $\Gamma=3$,
\begin{equation}
2S_0~\times ~S_{15} ~\times~ {\rm any\ number\ of\
}\{S_1,\cdots,S_{13}\}.\label{Example1}
\end{equation}
Eq. (\ref{Example1}) contains two $U_2$ fields and one $T_1^0$
field. With one $T_1^0$, however, we cannot satisfy the modular
invariance condition, Eq. (\ref{modinvk}), since all fields in
$\{S_1,\cdots,S_{13}\}$ are even twisted. So, the form
(\ref{Example1}) is not allowed. A similar conclusion is drawn if
we replace $S_{15}$ by $S_{23}$ in Eq. (\ref{Example1}). Even if
$\langle S_{15}\rangle\ne 0$ and $\langle S_{23}\rangle\ne 0$,
therefore, the coupling $u^cd^cd^c$ is not generated to all
orders.

Actually, there is a simpler argument for the absence of dimension
4 operators such as $u^cd^cd^c$. It comes from the
U(1)$_{\Gamma'}$ conservation. $u^cd^cd^c$ (and also $QLd^c$,
$LLe^c$) carries $\Gamma'=-1$, and the neutral singlets having
VEVs do not carry $S$ with positive $\Gamma'$. So, $u^cd^cd^c$ is
forbidden to all orders.

However by  $\langle S_{15}\rangle\ne 0$ and  $\langle
S_{23}\rangle\ne 0$, $d^c$ and $\overline{D}(T_4^0,T_6)$ can mix.
Eventually, this kind of mixing violates the $R$-parity. But the
violation will be suppressed by
$$
O\left(\frac{m_b}{m_D}\right)\sim 10^{-16}.
$$
A similar analysis can be done for $\Delta B=0,\Delta L\ne 0$ and
$R$ conserving operator $\overline{D}(T_4^0,T_6)QL$. Since proton
decay with dimesion 4 operators needs both of $u^cd^cd^c$ and
$\overline{D}(T_4^0,T_6)QL$, we will have the following suppression
factor for proton decay operator,
\begin{equation}
O\left(\frac{m_b}{m_D}\right)^2\sim 10^{-32}
\end{equation}
which is completely negligible. Then, proton decay proceeds
dominantly by the dimension 5 operators \cite{SakYan}. Being an SSM,
gauge boson exchanges do not lead to proton decay. But it is not
clear whether $p\to e^+ +(K,\overline{K})^0$ dominates over $p\to
e^++\pi^0$ since there is no reason that $d=5$ non-renormalizable
couplings are flavor distinguished.

\subsection{Effective $R$ parity of light particles and CDM candidate}
\label{subsec:Rlight}

The observation that the modular invariance condition removes the
coupling of the form (\ref{Example1}) hints that there might be an
effective $R$-parity among light (electroweak scale) particles. It
arises from the fact that the odd $R$ singlets of Table
\ref{tb:Singneutral} are in odd twisted sectors, and we need odd
number of these odd twisted sector VEVs to have $R$-parity
violating couplings. But the odd number of twisted sectors cannot
make modular invariant Yukawa couplings since the other
non-vanishing VEVs are carried by the fields in the even twisted
sectors.

$\nu^c$ in Eq. (\ref{eq:matter}) can obtain a large mass by
singlet VEVs, and considered to be in the intermediate scale. We
consider $H_u(U_2)$ and $H_d(U_2)$ are the electroweak scale Higgs
doublets. All the other vectorlike pairs in Table
\ref{tb:standardch} are considered to be at the string scale.
Thus, the light particles of Table \ref{tb:standardch} are $Q,
d^c, u^c, L, e^c$ of Eq. (\ref{eq:matter}), which carry $P=1$. If
we assume that boxed fields in Table \ref{tb:Singneutral} are
superheavy, the light (electroweak scale) Higgs fields, including
neutral singlets, carry even $P$ quantum numbers. In this way, we
have an effective $R$-parity among light fields. But the original
theory does not respect the $R$-parity, including all particles.
However, this $R$-parity violation must include heavy particles at
the string scale, which is not phenomenologically harmful. Since
any $R$-parity violation among light particles must occur at least
with a suppression factor of $O(M_{\rm string})$ for $\Delta B=0$
and $\Delta L\ne 0$ operators, the lightest supersymmetric
particle (LSP) defined among light fields must live at least
$10^{22}$ years, estimated by multiplying $(m_{\rm LSP}/m_p)^5$ to
the proton lifetime estimate obtained from dimension 5 operators.
Therefore, even though the $R$-parity is not exact, we have a cold
dark matter (CDM) candidate LSP which lives sufficiently long
enough.

\section{Model without exotics}\label{sec:ModelS}

The VEVs given in Eq. (\ref{VEVsinglets}) break U(1) gauge
symmetries with leaving only (SM gauge
gourp)${\rm\times[SO(10)\times U(1)_6]'}$.   Because the SM fields
are completely blind to U(1)$_6'$, it is possible to break a linear
combination of U(1)$_{\rm em}$ and U(1)$_6'$, leaving only one U(1)
unbroken. Let us call this unbroken U(1) the U(1) of quantum
electrodynamics, $\tilde{\rm U}(1)_{\rm em}$. We choose the symmetry
breaking direction such that there does not appear any exotics, i.e.
$\tilde{\rm U}(1)_{\rm em}$ charges of particles are integers for
color singlets, $+\frac23,-\frac13$ for color triplets (${\bf 3}$),
and $-\frac23,+\frac13$ for color anti-triplets (${\bf 3}^*$). The
electroweak hypercharge direction (\ref{ModelS}) fulfils this
possibility.

This is achieved by giving a VEV(s) to an exotic singlet(s). For
instance, let us choose just $\eta_1$ and $\overline{\eta}_6$. Both
$\langle \eta_1\overline{\eta}_6\rangle=0$ (Model E) and $\langle
\eta_1\overline{\eta}_6\rangle\neq 0$ (Model S) can be consistent
with SUSY, because the superpotential allows
$W=\eta_1\overline{\eta}_6S_4S_{10}
+(\eta_1\overline{\eta}_6S_4S_{10})^{13}+\cdots$, and both vacua can
satisfy the $F$- and $D$-flatness conditions.  If $\langle
\eta_1\overline{\eta}_6\rangle\neq 0$, the surviving U(1) gauge
symmetry is a linear combination of U(1)$_Y$ and U(1)$_6$, i.e. Eq.
(\ref{ModelS})
\begin{equation}
 \tilde Y= Y +\frac12 Q_6.
\end{equation}
Under this new U(1)$_{\tilde Y}$, all the exotics in  Model E carry
the regular hypercharges observed in the SSM.  With the new
U(1)$_{\tilde Y}$, thus, {\it all the exotics found in Model E are
moved into states with the standard charges} as shown in Table
\ref{tb:chargeModelS}. They still form vectorlike representations
under the SM gauge symmetry. Their mass terms discussed in Sec.
\ref{sec:Vecexotics} are still valid.  On the other hand, the
regularly charged states in  Model E, which originate from $U$,
$T_3$, $T_6$, $T_9$ and $T_k^0$ ($k=1,2,4,7$) sectors, are not
affected by this addition since they were not charged under U(1)$_6$
in the beginning.
\begin{table}[t]
\begin{center}
\begin{tabular}{|c|c|c|}
\hline &&\\[-1.2em]  $\tilde Y$ from (\ref{ModelS}) &
SU(3)$_c\times$SU(2)$_L$ & Exotics in Model E
\\[0.2em] \hline
$-\frac13$ & $({\bf 3},{\bf 1})$ & $\alpha_1^0$, ~$\alpha_3^0$,
~$2\cdot\alpha_4^0$
\\
$+\frac13$ & $({\bf 3^*},{\bf 1})$ & $\overline{\alpha}_2^0$,
~$3\cdot\overline{\alpha}_5^0$
\\
$-\frac23$ & $({\bf 3}^*,{\bf 1})$ &
$2\cdot\overline{\alpha}_6^{-1/3}$
\\
$+\frac23$ & $({\bf 3},{\bf 1})$ & $2\cdot{\alpha}_7^{1/3}$
\\ [0.2em]
\hline
 &&\\ [-1.2em]
 $-\frac12$ & $({\bf 1},{\bf 2})$ &  $\overline{\delta}_1$,
~$\delta_3$, ~$3\cdot\overline{\delta}_4$,
~$2\cdot\overline{\delta}_5$
\\
$\frac12$ & $({\bf 1},{\bf 2})$ &  $\delta_2$,~ $2\cdot\delta_6$,
~$2\cdot\delta_7$,~$\overline{\delta}_8$, ~$\delta_9$
\\ [0.2em]\hline
 & $$ &  $\xi_1$, $\eta_1$, $\eta_2$, $\eta_{4,5}$, $\overline{\eta}_6$,
$\overline{\eta}_7$, $\overline{\eta}_{9,10}$,
$3\cdot\overline{\eta}_{11}$, $2\cdot\overline{\eta}_{12}$,
\\
$0$ & $({\bf 1},{\bf 1})$ &  $2\cdot\overline{\eta}_{14,15,16}$,
$3\cdot\eta_{17}$, $3\cdot\eta_{18}$, $2\cdot\eta_{20,21,22}$,
\\
& $$ &  $\overline{\eta}_{23}$, $\overline{\eta}_{24}$,
$\overline{\eta}_{25}$, $\overline{\xi}_9$,
$\overline{\eta}_{27}$, $\overline{\eta}_{29}$, $\eta_{31}$,
$\eta_{33}$, $\eta_{34,35}$
\\ [0.2em]\hline
&&\\ [-1.2em] $-1$ & $({\bf 1},{\bf 1})$ &  $\eta_3$,
~$\overline{\xi}_3$, ~$2\cdot\overline{\xi}_4$,
~$2\cdot\overline{\xi}_5$, ~$2\cdot\eta_{19}$, ~$\overline{\xi}_8$,
~$\eta_{26}$, ~$\eta_{32}$
\\
$+1$ & $({\bf 1},{\bf 1})$ &  $\xi_2$, ~$\overline{\eta}_8$,
~$2\cdot\overline{\eta}_{13}$, ~$2\cdot\xi_6$, ~$2\cdot\xi_7$,
~$\overline{\eta}_{28}$, ~$\overline{\eta}_{30}$, ~$\xi_{10}$
\\[0.2em]
\hline
\end{tabular}
\end{center}
\caption{Model S contains no exotics. Previous exotics carry the
standard charges as shown in the first column. The charges of the
remaining states in Model S are the same as those in Model
E.}\label{tb:chargeModelS}
\end{table}
As mentioned below Eq. (\ref{ModelS}), the hypercharge operator in
Model S gives ${\rm sin}^2\theta_W^0=\frac{3}{14}$ at the string
scale. In this case, therefore, more (vectorlike) SU(3)$_c$
triplets and SU(2)$_L$ doublets at intermediate mass scales would
be needed to explain ${\rm sin}^2\theta_W\approx 0.23$ at the
electroweak scale. The discussion on the effective $R$-parity is
similar to that of Model E.

In this short section, we observed that models without exotics are
possible, but in such models it might be difficult to obtain
$\sin^2\theta_W^0= \frac38$ at the string scale.

\section{Conclusion}\label{sec:Conclusion}

We have constructed an SSM from a $\Z_{12-I}$ orbifold
compactification. In the vacuum chosen in (\ref{VEVsinglets}), we
achieve
\begin{itemize}
\item An SSM with three families with the third family in the
untwisted sector,
 \item  At the string scale,  ${\rm
sin}^2\theta_W^0=\frac38$,
 \item It is plausible to have one pair of
light Higgs doublets $H_u$ and $H_d$ from the untwisted sector,
 \item There exist Yukawa couplings for phenomenologically
satisfactory quark and lepton masses,
 \item All vectorlike color
triplets $D$ and $\overline{D}$ obtain masses, \item All exotic
particles are vectorlike and obtain masses, \item $D$- and
$F$-flat directions are possible,
 \item An effective $R$-parity (more accurately an effective
 matter parity), $P$, can be embedded
as a discrete group of gauged U(1)$_\Gamma$,
 \item All exotics
carry nonzero U(1)$_6$ quantum numbers,
 \item U(1)$_{\rm em}$ and
U(1)$_6$ are not broken. Therefore, there exist at least two
massless color singlet gauge bosons: photon and {\it exphoton}
(meaning the massless gauge boson coupling to exotic particles
only).
 \item If U(1)$_{\rm em}$ and U(1)$_6$ are properly broken to
give $\tilde{\rm U}(1)_{\rm em}$ unbroken, then one can convert
all exotics into states with the standard charges.
\end{itemize}

In sum we have shown that there exists a very satisfactory string
vacuum which meets all phenomenological constraints. At the least,
this paper shows the existence proof of the MSSM from superstring.
But why the VEVs of Eq. (\ref{VEVsinglets}) should be taken as
given there is not understood yet in this paper.

\acknowledgments{ This work is supported in part by the KRF ABRL
Grant No. R14-2003-012-01001-0. J.E.K. is also supported in part by
the KRF grants, No. R02-2004-000-10149-0 and  No.
KRF-2005-084-C00001.
 }

\appendix

\section{Massless Spectrum}\label{App:A}

The model presented in Eq. (\ref{Z12Imodel}) gives

\begin{eqnarray}
V^2-\phi^2 = 1, \quad a_3^2=4, \quad V\cdot a_3= 1,
\end{eqnarray}
\begin{eqnarray}
V_+^2-\phi^2 = 7, \quad V_-^2-\phi^2 = 3.
\end{eqnarray}
 Then, the gauge group is
\begin{eqnarray}
{\rm \left[\{SU(3)_c\times SU(2)_L\times U(1)_Y\}\times
U(1)_{B-L}\times U(1)^3\right]\times\left[SO(10)\times
U(1)^3\right]'}.
\end{eqnarray}
\begin{table} 
\begin{center}
\begin{tabular}{|c|c|c|c|}
\hline $P\cdot V$ & Visible States & $~\chi~$ & ~SM~
\\
\hline $\frac{7}{12}$ & $(\underline{++-};\underline{+-};+++)$ & L
& $Q$
\\
($U_1$) & $(---;\underline{+-};+--)$ & L & $L$
\\
\hline
$$ & $(0,0,0;\underline{1,0};0,0,1)$ &
L & $H_d$
\\
~~$\frac{4}{12}$~ ($U_2$)~ & $(0,0,0;\underline{-1,0};-1,0,0)$ & L &
$H_u$
\\
$$ & $(0,0,0;0,0;1,0,-1)$ & L & ${\bf
1_0}\equiv S_0$
\\
\hline
$$ & $(\underline{+--};--;+++)$ & L & $d^c$
\\
$\frac{1}{12}$ & $(+++;++;+++)$ & L &$\nu^c$
\\
($U_3$) & $(\underline{+--};++;+--)$ & L & $u^c$
\\
$$  & $(+++;--;-+-)$ & L & $e^c$
\\
\hline
\end{tabular}
\end{center}
\caption{Visible sector chiral fields from the $U$ sector. There is
no hidden sector chiral fields in the $U$ sector.
}\label{tb:untwistedvi}
\end{table}
\begin{table}[h]
\begin{center}
\begin{tabular}{|c|c|c|c|c||c|}
\hline  $P+6V$ & $~\chi~$ & $(N^L)_j$ & $\Theta_0$ & ${\cal P}_6$ &
SM
\\
\hline
 $\left(\underline{1,0,0};0,0;0^3\right)
 \left(\frac{-1}{2},\frac{1}{2};0^6\right)'$ &
L & $0$ & $\frac{-1}{3}$ & $3$& $3\cdot \overline{D}^{1/3}$
\\
 $\left(\underline{-1,0,0};0,0;0^3\right)
 \left(\frac{1}{2},\frac{-1}{2};0^6\right)'$ &
L & $0$ & $\frac{-1}{3}$ & $3$ & ~$3\cdot D^{-1/3}$~
\\
$\left(0,0,0;\underline{1,0};0^3\right)
\left(\frac{1}{2},\frac{-1}{2};0^6\right)'$
& L &  $0$ & $\frac{1}{6}$ & $2$ & $2\cdot H_d$
\\
 $\left(0,0,0;\underline{-1,0};0^3;0^3\right)
 \left(\frac{-1}{2},\frac{1}{2};0^6\right)'$ & L & $0$ &
 $\frac{1}{6}$ & $2$ & $2\cdot H_u$
\\
\hline
$\left(0^5;1,0,0\right)\left(\frac{-1}{2},\frac{1}{2};0^6\right)'$ &
L & $0$ & $\frac{-1}{6}$ & $2$ & $2\cdot {\bf 1_0}$
\\
 $\left(0^5;-1,0,0\right)\left(\frac{1}{2},\frac{-1}{2};0^6\right)'$
  & L & $0$ &
$\frac{1}{2}$ & $2$ & $2\cdot {\bf 1_0}$
\\
 $\left(0^5;0,0,1\right)\left(\frac{-1}{2},\frac{1}{2};0^6\right)'$
  & L & $0$ &
$\frac{1}{2}$ & $2$ & $2\cdot {\bf 1_0}$
\\
 $\left(0^5;0,0,-1\right)\left(\frac{1}{2},\frac{-1}{2};0^6\right)'$
  & L & $0$ &
$\frac{-1}{6}$ & $2$ & $2\cdot {\bf 1_0}$
\\
\hline
\end{tabular}
\end{center}
\caption{Massless states satisfying $(P+6V)\cdot W=0 $ mod $Z$ in
$T_6$.} \label{T6states}
\end{table}
\begin{table}[h]
\begin{center}
\begin{tabular}{|c|c|c|c|c|}
\hline  $P+3V$ & $~\chi~$ & $(N^L)_j$ & $\Theta_{L,R}$ & ~SM
[$SO(10)'$]~
\\
\hline
 $\left(\underline{\frac{3}{4},\frac{-1}{4},\frac{-1}{4}};
 \frac{-1}{4},\frac{-1}{4};
 \frac{1}{4},\frac{1}{4},\frac{1}{4}\right)
 \left(\frac{3}{4},\frac{1}{4},0;0^5\right)'$ &
L & $0$ & $\frac{1}{3}$ &  $\overline{D}^{1/3}$
\\
 $\left(\underline{\frac{3}{4},\frac{-1}{4},\frac{-1}{4}};
 \frac{-1}{4},\frac{-1}{4};
 \frac{1}{4},\frac{1}{4},\frac{1}{4}\right)
 \left(\frac{3}{4},\frac{1}{4},0;0^5\right)'$ &
R & $0$ & $0$ &  $2\cdot D^{-1/3~*}$, or\\
 &L&&&($2\cdot D^{-1/3}$ from
$T_9$)
\\
$\left(\frac{-1}{4},\frac{-1}{4},\frac{-1}{4};
\underline{\frac{3}{4},\frac{-1}{4}};
 \frac{1}{4},\frac{1}{4},\frac{1}{4}\right)
 \left(\frac{-1}{4},\frac{-3}{4},0;0^5\right)'$ &
L & $0$ & $\frac{1}{3}$ &  $H_d$
\\
 $\left(\frac{-1}{4},\frac{-1}{4},\frac{-1}{4};
 \underline{\frac{3}{4},\frac{-1}{4}};
 \frac{1}{4},\frac{1}{4},\frac{1}{4}\right)
 \left(\frac{-1}{4},\frac{-3}{4},0;0^5\right)'$ &
R &$0$ & $0$ & $2\cdot H_u^*$, or\\
&L&&& ($2\cdot H_u$ from $T_9$)
\\
\hline
$\left(\frac{-1}{4},\frac{-1}{4},\frac{-1}{4};\frac{-1}{4},\frac{-1}{4};
 \frac{-3}{4},\frac{1}{4},\frac{1}{4}\right)
 \left(\frac{-1}{4},\frac{-3}{4},0;0^5\right)'$ &
L & $0$ & $\frac{-1}{3}$ &  ${\bf 1_0}$
\\
 $\left(\frac{-1}{4},\frac{-1}{4},\frac{-1}{4};\frac{-1}{4},\frac{-1}{4};
 \frac{-3}{4},\frac{1}{4},\frac{1}{4}\right)
 \left(\frac{-1}{4},\frac{-3}{4},0;0^5\right)'$ &
R & $0$ & $\frac{1}{3}$ &  ${\bf 1_0}^*$
\\
 $\left(\frac{-1}{4},\frac{-1}{4},\frac{-1}{4};\frac{-1}{4},\frac{-1}{4};
 \frac{1}{4},\frac{1}{4},\frac{-3}{4}\right)
 \left(\frac{-1}{4},\frac{-3}{4},0;0^5\right)'$ &
L & $0$ & $0$ & $2\cdot{\bf 1_0}$
\\
 $\left(\frac{-1}{4},\frac{-1}{4},\frac{-1}{4};\frac{-1}{4},\frac{-1}{4};
 \frac{1}{4},\frac{1}{4},\frac{-3}{4}\right)
 \left(\frac{-1}{4},\frac{-3}{4},0;0^5\right)'$ &
R & $0$ & $\frac{-1}{3}$ &  ${\bf 1_0}^*$
\\
 $\left(\frac{1}{4},\frac{1}{4},\frac{1}{4};\frac{1}{4},\frac{1}{4};
 \frac{-1}{4},\frac{-1}{4},\frac{-1}{4}\right)
 \left(\frac{3}{4},\frac{1}{4},0;0^5\right)'$ &
L & $1_3$ & $\frac{1}{3}$ &  ${\bf 1_0}$
\\
 $\left(\frac{1}{4},\frac{1}{4},\frac{1}{4};\frac{1}{4},\frac{1}{4};
 \frac{-1}{4},\frac{-1}{4},\frac{-1}{4}\right)
 \left(\frac{3}{4},\frac{1}{4},0;0^5\right)'$ &
R & $1_3$ & $0$ &  $2\cdot{\bf 1_0}^*$
\\
\hline
 $\left(\frac{1}{4},\frac{1}{4},\frac{1}{4};\frac{1}{4},\frac{1}{4};
 \frac{-1}{4},\frac{-1}{4},\frac{-1}{4}\right)
 \left(\frac{-1}{4},\frac{1}{4},0;\underline{\pm 1,0^4}\right)'$ &
L & $0$ & $0$ &  $2\cdot{\bf 1_0}~[{\bf 10'}]$
\\
 $\left(\frac{1}{4},\frac{1}{4},\frac{1}{4};\frac{1}{4},\frac{1}{4};
 \frac{-1}{4},\frac{-1}{4},\frac{-1}{4}\right)
 \left(\frac{-1}{4},\frac{1}{4},0;\underline{\pm 1,0^4}\right)'$ &
R & $0$ & $\frac{-1}{3}$ &  ${\bf 1_0}^*~[{\bf 10'}]$
\\
\hline
\end{tabular}
\end{center}
\caption{Massless states satisfying $(P+3V)\cdot W=0 $ mod $Z$ in
$T_3$. The starred chirality R fields in $T_3$ can be represented
also by un-starred chirality L fields with the opposite quantum
numbers in $T_9$, as shown in two lines. There are, in total, three
${\bf 10'}$s of the hidden $SO(10)'$ from the $T_3$ and $T_9$
sectors. The other states in $T_3$ and $T_9$ are singlets under the
hidden gauge group. The multiplicity is shown as the coefficient in
the last column.} \label{T3states}
\end{table}
\begin{table} 
\begin{center}
\begin{tabular}{|c|c|c|c||c|}
\hline  $P+2V$ & $~\chi~$ & $(N^L)_j$ & ${\cal P}_2(f_0)$ & SM
\\
\hline $\left(0^5;\frac{-2}{3},\frac{-2}{3},\frac{-1}{3}\right)
(\frac{1}{2},\frac{-1}{2};0^6)'$ & L & $0$ & $1$ & ${\bf 1_0}$
\\
$\left(0^5;\frac{-2}{3},\frac{1}{3},\frac{2}{3}\right)
(\frac{-1}{2},\frac{1}{2};0^6)'$ & L &$0$ & $1$ & ${\bf 1_0}$
\\
$\left(0^5;\frac{1}{3},\frac{-2}{3},\frac{2}{3}\right)
(\frac{-1}{2},\frac{1}{2};0^6)'$ & L & $0$ & $1$ & ${\bf 1_0}$
 \\
$\left(0^5;\frac{1}{3},\frac{1}{3},\frac{-1}{3}\right)
(\frac{1}{2},\frac{-1}{2};0^6)'$ & L & $2_{\bar{1}},~2_3$ & $1+1$
& $2\cdot {\bf 1_0}$
\\
$\left(0^5;\frac{1}{3},\frac{1}{3},\frac{-1}{3}\right)
(\frac{-1}{2},\frac{1}{2};0^6)'$ & L &
$1_{\bar{2}},~\{1_{\bar{1}}+1_3\}$ & $1+1$ & ~$2\cdot {\bf 1_0}$~
\\
\hline \hline
 $P+2V_+$ & $~\chi~$ & $(N^L)_j$ & ${\cal P}_2(f_+)$&
~SM~
\\
\hline
$\left(\underline{\frac{5}{6},\frac{-1}{6},\frac{-1}{6}};
\frac{1}{6},\frac{1}{6};
\frac{1}{6},\frac{-1}{6},\frac{-1}{2}\right)
\left(\frac{1}{2},\frac{-1}{6},\frac{1}{3};0^5\right)'$ & L & $0$ &
$1$ & $\overline{\alpha}^0$
\\
$\left(\frac{1}{3},\frac{1}{3},\frac{1}{3};\frac{-1}{3},\frac{-1}{3};
\frac{-1}{3},\frac{1}{3},0\right)
\left(\frac{-1}{2},\frac{-1}{6},\frac{-2}{3};0^5\right)'$ & L & $0$
& $1$ & $\xi^{2/3}$
\\
$\left(\frac{1}{3},\frac{1}{3},\frac{1}{3};\frac{-1}{3},\frac{-1}{3};
\frac{-1}{3},\frac{1}{3},0\right)
\left(\frac{1}{2},\frac{-1}{6},\frac{1}{3};0^5\right)'$ & L &
$1_3$ & $1$ & $\xi^{2/3}$
\\
$\left(\frac{-1}{6},\frac{-1}{6},\frac{-1}{6};\frac{1}{6},\frac{1}{6};
\frac{-5}{6},\frac{-1}{6},\frac{-1}{2}\right)
\left(\frac{1}{2},\frac{-1}{6},\frac{1}{3};0^5\right)'$ & L & $0$ &
$1$ & $\eta^{-1/3}$
\\
$\left(\frac{-1}{6},\frac{-1}{6},\frac{-1}{6};\frac{1}{6},\frac{1}{6};
\frac{1}{6},\frac{-1}{6},\frac{1}{2}\right)
\left(\frac{-1}{2},\frac{5}{6},\frac{1}{3};0^5\right)'$ & L & $0$ &
$1$ & $\eta^{-1/3}$
\\
$\left(\frac{-1}{6},\frac{-1}{6},\frac{-1}{6};\frac{1}{6},\frac{1}{6};
\frac{1}{6},\frac{-1}{6},\frac{1}{2}\right)
\left(\frac{-1}{2},\frac{-1}{6},\frac{-2}{3};0^5\right)'$ & L &
$1_3$ & $1$ & $\eta^{-1/3}$
\\
$\left(\frac{-1}{6},\frac{-1}{6},\frac{-1}{6};\frac{1}{6},\frac{1}{6};
\frac{1}{6},\frac{-1}{6},\frac{1}{2}\right)
\left(\frac{1}{2},\frac{-1}{6},\frac{1}{3};0^5\right)'$ & L &
$2_{\bar{1}},~ 2_3$ & $1+1$ & $~2\cdot \eta^{-1/3}~$
\\
\hline \hline
 $P+2V_-$ & $~\chi~$ & $(N^L)_j$ & ${\cal P}_2(f_-)$&
~SM~
\\
\hline
$\left(\underline{\frac{-5}{6},\frac{1}{6},\frac{1}{6}};
\frac{-1}{6},\frac{-1}{6};
\frac{-1}{2},\frac{-1}{6},\frac{-1}{6}\right)
\left(\frac{-1}{2},\frac{1}{6},\frac{-1}{3};0^5\right)'$ & L & $0$ &
$1$ & $\alpha^0$
\\
$\left(\frac{-1}{3},\frac{-1}{3},\frac{-1}{3};\frac{1}{3},\frac{1}{3};
0,\frac{1}{3},\frac{1}{3}\right)
\left(\frac{-1}{2},\frac{1}{6},\frac{-1}{3};0^5\right)'$ & L &
$1_3$ & $1$ & $\overline{\xi}^{-2/3}$
\\
$\left(\frac{1}{6},\frac{1}{6},\frac{1}{6};\frac{-1}{6},\frac{-1}{6};
\frac{-1}{2},\frac{-1}{6},\frac{5}{6}\right)
\left(\frac{-1}{2},\frac{1}{6},\frac{-1}{3};0^5\right)'$ & L & $0$ &
$1$ & $\overline{\eta}^{1/3}$
\\
$\left(\frac{1}{6},\frac{1}{6},\frac{1}{6};\frac{-1}{6},\frac{-1}{6};
\frac{-1}{2},\frac{5}{6},\frac{-1}{6}\right)
\left(\frac{-1}{2},\frac{1}{6},\frac{-1}{3};0^5\right)'$ & L & $0$ &
$1$ & $\overline{\eta}^{1/3}$
\\
$\left(\frac{1}{6},\frac{1}{6},\frac{1}{6};\frac{-1}{6},\frac{-1}{6};
\frac{1}{2},\frac{-1}{6},\frac{-1}{6}\right)
\left(\frac{1}{2},\frac{1}{6},\frac{2}{3};0^5\right)'$ & L & $1_3$
& $1$ & $\overline{\eta}^{1/3}$
\\
$\left(\frac{1}{6},\frac{1}{6},\frac{1}{6};\frac{-1}{6},\frac{-1}{6};
\frac{1}{2},\frac{-1}{6},\frac{-1}{6}\right)
\left(\frac{-1}{2},\frac{1}{6},\frac{-1}{3};0^5\right)'$ & L &
$1_{\bar{2}},~ \{1_{\bar{1}}+1_3\}$ & $1+1$ & $~2\cdot
\overline{\eta}^{1/3}~$
\\
\hline
\end{tabular}
\end{center}
\caption{Chiral matter fields satisfying $\Theta_0=0$ in the $T_2^0$
sector,  $\Theta_+=0$ in the $T_{2}^+$ sector, and $\Theta_-=0$ in
the $T_{2}^-$ sector.}\label{tb:T2}
\end{table}
In this model, there are eight U(1) symmetries whose charges are
\begin{eqnarray}
&& Y=\textstyle\left(\frac{1}{3}, \frac{1}{3}, \frac{1}{3};
\frac{-1}{2}, \frac{-1}{2}; 0^3\right)\left(0^8\right)'
\\
&& {B-L}=\textstyle \left(\frac{2}{3}, \frac{2}{3}, \frac{2}{3};
0^2; 0^3\right)\left(0^8\right)'
\\
&& Q_1=\left(0^5;2, 0, 0\right)\left(0^8\right)'
\\
&& Q_2=\left(0^5; 0, 2, 0\right)\left(0^8\right)'
\\
&& Q_3=\left(0^5; 0, 0, 2\right)\left(0^8\right)'
\\
&& Q_4=\left(0^8\right)\left(2, 0, 0; 0^5\right)'
\\
&& Q_5=\left(0^8\right)\left(0, 2, 0; 0^5\right)'
\\
&& Q_6=\left(0^8\right)\left(0, 0, 2; 0^5\right)' .
\end{eqnarray}
There are two familiar U(1) charges
\begin{eqnarray}
Y=\textstyle\left(\frac{1}{3},\frac{1}{3},\frac{1}{3};
\frac{-1}{2},\frac{-1}{2}; 0,0,0\right)\left(0^8\right)',
\end{eqnarray}
\begin{eqnarray}
Q_{B-L}\equiv
 B-L=\textstyle\left(\frac{2}{3},\frac{2}{3},\frac{2}{3};0,0;
0,0,0\right)\left(0^8\right)'.
\end{eqnarray}
Note that $X$ of the flipped SU(5) is a combination of $B-L$ and
$Y$,
\begin{equation}
X=\textstyle\left(-2,-2,-2,-2,-2,;
0,0,0\right)\left(0^8\right)'=-5(B-L)+4Y.
\end{equation}

The U(1)$_\Gamma$ charge used in the text is
\begin{equation}
\Gamma=\textstyle X+\frac14(Q_4+Q_5) -(Q_2+Q_3) +6(B-L).
\end{equation}

Using the technique and notation of \cite{KimKyaegut}, massless
fields are calculated. In Table \ref{tb:untwistedvi}, we list the
massless fields from the untwisted sector. There is one singlet
$S_0$ which cannot be a member of the SO(10) spinor. In Tables
\ref{T6states} and \ref{T3states}, we list massless fields in
$T_6$ and $T_3({\rm and}\ T_9)$ which are not affected by Wilson
lines. In Tables \ref{tb:T2}, \ref{tb:T4}, \ref{tb:T10}, and
\ref{tb:T5}, we list massless fields of $T_2, T_4, T_1,$ and $T_5$
sectors, respectively. For the SM particles, we use the familiar
notations: $Q, u^c, d^c, L, e^c, \nu^c$ for sixteen fields of the
SM and $S$ for ${\rm SO(10)}'$ singlets. The Higgs doublets are
denoted by $H_u$ and $H_d$. The color triplets with
\Qem$=-\frac13$, which in principle can mix with $d$, are denoted
as $D$.

Exotic particles appear in the sectors affected by Wilson lines:
$T_2^{\pm}, T_4^{\pm}, T_1^{\pm},$ and $T_5^{\pm}$. For these
exotics, we use the following notations:
\begin{equation}
\begin{array}{ll}
\alpha_i,\  \overline{\alpha}_j:  &{\rm color\ exotics\
}{\bf 3}\ {\rm and\ }{\bf 3}^*\\
\delta_i,\  \overline{\delta}_j: & {\rm SU(2)\ doublet\ exotics}\\
\xi_i,\ \overline{\xi}_j: & Q_{\rm em}=\pm\frac23\ {\rm SU(3)\times
SU(2)\  singlet\ exotics}\\
\eta_i,\ \overline{\eta}_j: &Q_{\rm em}=\mp\frac13\ {\rm SU(3)\times
SU(2)\  singlet\ exotics}
\end{array}
\end{equation}
If some exotics do not obtain mass, the model must be excluded from
phenomenological consideration. In the text, we have shown that all
exotics obtain masses. This {\it massive exotics condition}
determines the vacuum where nonvanishing VEVs of $S$ fields are
dictated. There are many possibilities for giving masses to exotic
particles. In this paper, we chose the minimum number of neutral
singlet VEVs, Eq. (\ref{VEVsinglets}).


\begin{table}[t]
{\small
\begin{center}
\begin{tabular}{|c|c|c|c|c||c|}
\hline  $P+4V$ & $~\chi~$ & $(N^L)_j$ &
 $\Theta_0$ & ${\cal P}_4(f_0)$ & SM
\\
\hline
$\left(\underline{+--};--;\frac{1}{6},\frac{1}{6},\frac{-1}{6}\right)
(0^8)'$ & L & $0$ & $\frac{-1}{4}$ & $2$ &$2\cdot d^c$
\\
$\left(---;\underline{+-};\frac{1}{6},\frac{1}{6},\frac{-1}{6}\right)
(0^8)'$ & L & $0$ & $\frac{-1}{4}$ & $2$ & $2\cdot L$
\\
$\left(\underline{+--}++;\frac{1}{6},\frac{1}{6},\frac{-1}{6}\right)
(0^8)'$ & L & $0$ & $\frac{1}{4}$ & $2$ & $2\cdot u^c$
\\
$\left(\underline{++-};\underline{+-};
\frac{1}{6},\frac{1}{6},\frac{-1}{6}\right)
(0^8)'$ & L & $0$ & $\frac{1}{4}$ & $2$ & $2\cdot Q$
\\
$\left(+++;--;\frac{1}{6},\frac{1}{6},\frac{-1}{6}\right) (0^8)'$ &
L & $0$ & $\frac{1}{4}$ & $2$ & $2\cdot e^c$
\\
$\left(+++;++;\frac{1}{6},\frac{1}{6},\frac{-1}{6}\right) (0^8)'$ &
L & $0$ & $\frac{-1}{4}$ & $2$ & $2\cdot \nu^c$
\\
$\left(\underline{1,0,0};0,0;\frac{-1}{3},\frac{-1}{3},
\frac{1}{3}\right)(0^8)'$
& L & $0$ & $0$ & $3$ & $3\cdot \overline{D}^{1/3}$
\\
$\left(\underline{-1,0,0};0,0;\frac{-1}{3},\frac{-1}{3},
\frac{1}{3}\right)(0^8)'$
& L & $0$ & $\frac{1}{2}$ & $2$ & ~$2\cdot D^{-1/3}$~
\\
$\left(0,0,0;\underline{1,0};\frac{-1}{3},\frac{-1}{3},
\frac{1}{3}\right)(0^8)'$
& L & $0$ & $0$ & $3$ & $3\cdot H_d$
\\
$\left(0,0,0;\underline{-1,0};\frac{-1}{3},\frac{-1}{3},
\frac{1}{3}\right)(0^8)'$
& L & $0$ & $\frac{1}{2}$ & $2$ & $2\cdot H_u$
\\
$\left(0,0,0;0,0;\frac{2}{3},\frac{2}{3},\frac{-2}{3}\right)(0^8)'$
& L & $0$ & $\frac{1}{2}$ & $2$ & $2\cdot {\bf 1_0}$
\\
$\left(0,0,0;0,0;\frac{-1}{3},\frac{-1}{3},\frac{-2}{3}\right)(0^8)'$
& L & $1_{\bar{1}},1_2,1_3$ & $\frac{-1}{4},0,\frac{1}{4}$ &
$2+3+2$ & $7\cdot {\bf 1_0}$
\\
$\left(0,0,0;0,0;\frac{-1}{3},\frac{2}{3},\frac{1}{3}\right)(0^8)'$
& L & $1_{\bar{1}},1_2,1_3$ &
$\frac{1}{4},\frac{1}{2},\frac{-1}{4}$ & $2+2+2$ & $6\cdot {\bf
1_0}$
\\
$\left(0,0,0;0,0;\frac{2}{3},\frac{-1}{3},\frac{1}{3}\right)(0^8)'$
& L & $1_{\bar{1}},1_2,1_3$ &
$\frac{1}{4},\frac{1}{2},\frac{-1}{4}$ & $2+2+2$ & $6\cdot {\bf
1_0}$
\\
\hline \hline
 $P+4V_+$ & $~\chi~$ & $(N^L)_j$ & $\Theta_+$ & ${\cal
P}_4(f_+)$ & SM
\\
\hline $\left(\underline{\frac{-5}{6},\frac{1}{6},\frac{1}{6}};
\frac{-1}{6},\frac{-1}{6};
\frac{-1}{6},\frac{1}{6},\frac{-1}{2}\right)
\left(0,\frac{-1}{3},\frac{-1}{3};0^5\right)'$ & L & $0$ &
$\frac{1}{2}$ & $2$ & $2\cdot \alpha^0$
\\
$\left(\underline{\frac{2}{3},\frac{-1}{3},\frac{-1}{3}};
\frac{1}{3},\frac{1}{3}; \frac{1}{3},\frac{-1}{3},0\right)
\left(0,\frac{-1}{3},\frac{-1}{3};0^5\right)'$ & L & $0$ &
$\frac{1}{4}$ & $2$ & ~$2\cdot \overline{\alpha}^{-1/3}$~
\\
$\left(\frac{1}{6},\frac{1}{6},\frac{1}{6};
\underline{\frac{5}{6},\frac{-1}{6}};
\frac{-1}{6},\frac{1}{6},\frac{-1}{2}\right)
\left(0,\frac{-1}{3},\frac{-1}{3};0^5\right)'$ & L & $0$ & $0$ & $3$
& $3\cdot \overline{\delta}^{-1/6}$
\\
$\left(\frac{-1}{3},\frac{-1}{3},\frac{-1}{3};
\underline{\frac{-2}{3},\frac{1}{3}};
\frac{1}{3},\frac{-1}{3},0\right)
\left(0,\frac{-1}{3},\frac{-1}{3};0^5\right)'$ & L & $0$ &
$\frac{-1}{4}$ & $2$ & $2\cdot \overline{\delta}^{-1/6}$
\\
$\left(\frac{-1}{3},\frac{-1}{3},\frac{-1}{3};\frac{1}{3},\frac{1}{3};
\frac{-2}{3},\frac{-1}{3},0\right)
\left(0,\frac{-1}{3},\frac{-1}{3};0^5\right)'$ & L & $0$ &
$\frac{1}{4}$ & $2$ & $2\cdot \overline{\xi}^{-2/3}$
\\
$\left(\frac{-1}{3},\frac{-1}{3},\frac{-1}{3};\frac{1}{3},\frac{1}{3};
\frac{1}{3},\frac{2}{3},0\right)
\left(0,\frac{-1}{3},\frac{-1}{3};0^5\right)'$ & L & $0$ &
$\frac{-1}{4}$ & $2$ & $2\cdot \overline{\xi}^{-2/3}$
\\
$\left(\frac{1}{6},\frac{1}{6},\frac{1}{6};\frac{-1}{6},\frac{-1}{6};
\frac{-1}{6},\frac{-5}{6},\frac{-1}{2}\right)
\left(0,\frac{-1}{3},\frac{-1}{3};0^5\right)'$ & L & $0$ & $0$ & $3$
& $3\cdot \overline{\eta}^{1/3}$
\\
$\left(\frac{1}{6},\frac{1}{6},\frac{1}{6};\frac{-1}{6},\frac{-1}{6};
\frac{5}{6},\frac{1}{6},\frac{-1}{2}\right)
\left(0,\frac{-1}{3},\frac{-1}{3};0^5\right)'$ & L & $0$ &
$\frac{1}{2}$ & $2$ & $2\cdot \overline{\eta}^{1/3}$
\\
$\left(\frac{1}{6},\frac{1}{6},\frac{1}{6};\frac{-1}{6},\frac{-1}{6};
\frac{-1}{6},\frac{1}{6},\frac{1}{2}\right)
\left(0,\frac{2}{3},\frac{2}{3};0^5\right)'$ & L & $0$&
$\frac{1}{4}$ & $2$ & $2\cdot \overline{\eta}^{1/3}$
\\
$\left(\frac{1}{6},\frac{1}{6},\frac{1}{6};\frac{-1}{6},\frac{-1}{6};
\frac{-1}{6},\frac{1}{6},\frac{1}{2}\right)
\left(0,\frac{-1}{3},\frac{-1}{3};0^5\right)'$ & L & $1_{\bar{1}},
1_2, 1_3$ & $\frac{1}{4}, \frac{1}{2}, \frac{-1}{4}$ & $2+2+2$ &
$6\cdot \overline{\eta}^{1/3}$
\\
\hline \hline
  $P+4V_-$ & $~\chi~$ & $(N^L)_j$ & $\Theta_-$ & ${\cal
P}_4(f_-)$ & SM
\\
\hline $\left(\underline{\frac{5}{6},\frac{-1}{6},\frac{-1}{6}};
\frac{1}{6},\frac{1}{6}; \frac{-1}{2},\frac{1}{6},\frac{1}{6}\right)
\left(0,\frac{1}{3},\frac{1}{3};0^5\right)'$ & L & $0$ & $0$ & $3$ &
$3\cdot \overline{\alpha}^0$
\\
$\left(\underline{\frac{-2}{3},\frac{1}{3},\frac{1}{3}};
\frac{-1}{3},\frac{-1}{3}; 0,\frac{-1}{3},\frac{-1}{3}\right)
\left(0,\frac{1}{3},\frac{1}{3};0^5\right)'$ & L & $0$ &
$\frac{1}{4}$ & $2$ & $2\cdot \alpha^{1/3}$
\\
$\left(\frac{-1}{6},\frac{-1}{6},\frac{-1}{6};
\underline{\frac{-5}{6},\frac{1}{6}};
\frac{-1}{2},\frac{1}{6},\frac{1}{6}\right)
\left(0,\frac{1}{3},\frac{1}{3};0^5\right)'$ & L & $0$ &
$\frac{1}{2}$ & $2$ & $2\cdot \delta^{1/6}$
\\
$\left(\frac{1}{3},\frac{1}{3},\frac{1}{3};
\underline{\frac{2}{3},\frac{-1}{3}};
0,\frac{-1}{3},\frac{-1}{3}\right)
\left(0,\frac{1}{3},\frac{1}{3};0^5\right)'$ & L & $0$ &
$\frac{-1}{4}$ & $2$ & $2\cdot \delta^{1/6}$
\\
$\left(\frac{1}{3},\frac{1}{3},\frac{1}{3};\frac{-1}{3},\frac{-1}{3};
0,\frac{2}{3},\frac{-1}{3}\right)
\left(0,\frac{1}{3},\frac{1}{3};0^5\right)'$ & L & $0$ &
$\frac{1}{4}$ & $2$ & $2\cdot \xi^{2/3}$
\\
$\left(\frac{1}{3},\frac{1}{3},\frac{1}{3};\frac{-1}{3},\frac{-1}{3};
0,\frac{-1}{3},\frac{2}{3}\right)
\left(0,\frac{1}{3},\frac{1}{3};0^5\right)'$ & L & $0$ &
$\frac{1}{4}$ & $2$ & $2\cdot \xi^{2/3}$
\\
$\left(\frac{-1}{6},\frac{-1}{6},\frac{-1}{6};\frac{1}{6},\frac{1}{6};
\frac{-1}{2},\frac{-5}{6},\frac{1}{6}\right)
\left(0,\frac{1}{3},\frac{1}{3};0^5\right)'$ & L & $0$ & $0$ & $3$ &
$3\cdot {\eta}^{-1/3}$
\\
$\left(\frac{-1}{6},\frac{-1}{6},\frac{-1}{6};\frac{1}{6},\frac{1}{6};
\frac{-1}{2},\frac{1}{6},\frac{-5}{6}\right)
\left(0,\frac{1}{3},\frac{1}{3};0^5\right)'$ & L & $0$&  $0$ & $3$ &
$3\cdot {\eta}^{-1/3}$
\\
$\left(\frac{-1}{6},\frac{-1}{6},\frac{-1}{6};\frac{1}{6},\frac{1}{6};
\frac{1}{2},\frac{1}{6},\frac{1}{6}\right)
\left(0,\frac{-2}{3},\frac{-2}{3};0^5\right)'$ & L & $0$ &
$\frac{-1}{4}$ & $2$ & $2\cdot {\eta}^{-1/3}$
\\
$\left(\frac{-1}{6},\frac{-1}{6},\frac{-1}{6};\frac{1}{6},\frac{1}{6};
\frac{1}{2},\frac{1}{6},\frac{1}{6}\right)
\left(0,\frac{1}{3},\frac{1}{3};0^5\right)'$ & L & $1_{\bar{1}},
1_2, 1_3$ & $\frac{1}{4}, \frac{1}{2}, \frac{-1}{4}$ & $2+2+2$ &
~$6\cdot {\eta}^{-1/3}$~
\\
\hline
\end{tabular}
\end{center}
\caption{Chiral matter fields in the $T_4^0$, $T_4^+$, and $T_4^-$
sectors. }\label{tb:T4}
 }
\end{table}

\begin{table}[h]
\begin{center}
\begin{tabular}{|c|c|c|c||c|}
\hline  $P+V$ & $~\chi~$ & $(N^L)_j$ & ${\cal P}_1(f_0)$ & ~SM~
\\
\hline
 $\left(\frac{1}{4},\frac{1}{4},\frac{1}{4};\frac{1}{4},\frac{1}{4};
 \frac{5}{12},\frac{5}{12},\frac{1}{12}\right)
 (\frac{1}{4},\frac{3}{4};0^6)'$ & L & $1_3$ & $1$ & ${\bf 1_0}$
 \\
$\left(\frac{1}{4},\frac{1}{4},\frac{1}{4};\frac{1}{4},\frac{1}{4};
 \frac{5}{12},\frac{5}{12},\frac{1}{12}\right)
 (\frac{-3}{4},\frac{-1}{4};0^6)'$ & L & $1_3$ & $1$ & ${\bf 1_0}$
 \\
  $\left(\frac{-1}{4},\frac{-1}{4},\frac{-1}{4};
  \frac{-1}{4},\frac{-1}{4};
 \frac{-1}{12},\frac{-1}{12},\frac{-5}{12}\right)
 (\frac{1}{4},\frac{3}{4};0^6)'$ & L & $2_3$ & $1$ & ${\bf 1_0}$
 \\
$\left(\frac{-1}{4},\frac{-1}{4},\frac{-1}{4};\frac{-1}{4},\frac{-1}{4};
 \frac{-1}{12},\frac{-1}{12},\frac{-5}{12}\right)
 (\frac{-3}{4},\frac{-1}{4};0^6)'$ & L & $2_3$ & $1$ &${\bf 1_0}$
 \\
\hline \hline $P+V_+$ & $~\chi~$ & $(N^L)_j$ & ${\cal P}_1(f_+)$ &
SM
\\
\hline $\left(\underline{\frac{-7}{12},\frac{5}{12},\frac{5}{12}};
\frac{1}{12},\frac{1}{12};
\frac{-5}{12},\frac{-1}{12},\frac{3}{12}\right)
\left(\frac{3}{12},\frac{5}{12},\frac{-4}{12};0^5\right)'$ & L &
$1_3$ & $1$ & $\alpha^0$
\\
$\left(\frac{-1}{12},\frac{-1}{12},\frac{-1}{12};
\underline{\frac{7}{12},\frac{-5}{12}};
\frac{1}{12},\frac{5}{12},\frac{-3}{12}\right)
\left(\frac{3}{12},\frac{5}{12},\frac{-4}{12};0^5\right)'$ & L &
$2_3$ & $1$ & $\overline{\delta}^{-1/6}$
\\
$\left(\frac{-1}{12},\frac{-1}{12},\frac{-1}{12};
\frac{7}{12},\frac{7}{12};
\frac{1}{12},\frac{-7}{12},\frac{-3}{12}\right)
\left(\frac{3}{12},\frac{5}{12},\frac{-4}{12};0^5\right)'$ & L & $0$
& $1$ & $\overline{\xi}^{-2/3}$
\\
$\left(\frac{-1}{12},\frac{-1}{12},\frac{-1}{12};
\frac{-5}{12},\frac{-5}{12};
\frac{1}{12},\frac{5}{12},\frac{9}{12}\right)
\left(\frac{3}{12},\frac{5}{12},\frac{-4}{12};0^5\right)'$ & L & $0$
& $1$ & $\overline{\eta}^{1/3}$
\\
$\left(\frac{5}{12},\frac{5}{12},\frac{5}{12};
\frac{1}{12},\frac{1}{12};
\frac{7}{12},\frac{-1}{12},\frac{3}{12}\right)
\left(\frac{3}{12},\frac{5}{12},\frac{-4}{12};0^5\right)'$ & L &
$1_3$ & $1$ & $\overline{\eta}^{1/3}$
\\
$\left(\frac{-1}{12},\frac{-1}{12},\frac{-1}{12};
\frac{-5}{12},\frac{-5}{12};
\frac{1}{12},\frac{-7}{12},\frac{-3}{12}\right)
\left(\frac{3}{12},\frac{5}{12},\frac{-4}{12};0^5\right)'$ & L &
$2_3$ & $1$ & $\overline{\eta}^{1/3}$
\\
\hline \hline $P+V_-$ & $~\chi~$ & $(N^L)_j$ & ${\cal P}_1(f_-)$ &
SM
\\
\hline
 $\left(\frac{1}{12},\frac{1}{12},\frac{1}{12};
\underline{\frac{-7}{12},\frac{5}{12}};
\frac{3}{12},\frac{-1}{12},\frac{-1}{12}\right)
\left(\frac{-9}{12},\frac{1}{12},\frac{4}{12};0^5\right)'$ & L &
$1_3$ & $1$ & $\delta^{1/6}$
\\
$\left(\frac{1}{12},\frac{1}{12},\frac{1}{12};
\underline{\frac{-7}{12},\frac{5}{12}};
\frac{3}{12},\frac{-1}{12},\frac{-1}{12}\right)
\left(\frac{3}{12},\frac{1}{12},\frac{-8}{12};0^5\right)'$ & L &
$2_3$ & $1$ & $\delta^{1/6}$
\\
$\left(\frac{1}{12},\frac{1}{12},\frac{1}{12};
\frac{5}{12},\frac{5}{12};
\frac{-9}{12},\frac{-1}{12},\frac{-1}{12}\right)
\left(\frac{3}{12},\frac{1}{12},\frac{-8}{12};0^5\right)'$ & L & $0$
& $1$ & ~$\eta^{-1/3}$~
\\
\hline
\end{tabular}
\end{center}
\caption{Chiral matter fields satisfying $\Theta=0$ in the $T_1^0$
and $T_1^\pm$ sectors.}\label{tb:T10}
\end{table}


\begin{table}[h]
\begin{center}
\begin{tabular}{|c|c|c|c||c|}
\hline  $P+5V$ & $~\chi~$ & $(N^L)_j$ & ${\cal P}_5(f_0)$ & SM
\\
\hline
 $\left(\frac{-1}{4},\frac{-1}{4},\frac{-1}{4},\frac{-1}{4},\frac{-1}{4};
 \frac{7}{12},\frac{-5}{12},\frac{-1}{12}\right)
 (\frac{1}{4},\frac{3}{4};0^6)'$ & R & $0$ & $1$ & ${\bf 1_0}^*$
 \\
$\left(\frac{-1}{4},\frac{-1}{4},\frac{-1}{4},\frac{-1}{4},\frac{-1}{4};
 \frac{7}{12},\frac{-5}{12},\frac{-1}{12}\right)
 (\frac{-3}{4},\frac{-1}{4};0^6)'$ & R & $0$ & $1$ & ${\bf 1_0}^*$
 \\
  $\left(\frac{-1}{4},\frac{-1}{4},\frac{-1}{4},\frac{-1}{4},\frac{-1}{4};
 \frac{-5}{12},\frac{7}{12},\frac{-1}{12}\right)
 (\frac{1}{4},\frac{3}{4};0^6)'$ & R & $0$ & $1$ & ${\bf 1_0}^*$
 \\
  $\left(\frac{-1}{4},\frac{-1}{4},\frac{-1}{4},\frac{-1}{4},\frac{-1}{4};
 \frac{-5}{12},\frac{7}{12},\frac{-1}{12}\right)
 (\frac{-3}{4},\frac{-1}{4};0^6)'$ & R & $0$ & $1$ & ${\bf 1_0}^*$
 \\
$\left(\frac{1}{4},\frac{1}{4},\frac{1}{4},\frac{1}{4},\frac{1}{4};
 \frac{1}{12},\frac{1}{12},\frac{-7}{12}\right)
 (\frac{1}{4},\frac{3}{4};0^6)'$ & R & $1_1$ & $1$ &${\bf 1_0}^*$
 \\
  $\left(\frac{1}{4},\frac{1}{4},\frac{1}{4},\frac{1}{4},\frac{1}{4};
 \frac{1}{12},\frac{1}{12},\frac{-7}{12}\right)
 (\frac{-3}{4},\frac{-1}{4};0^6)'$ & R & $1_1$ & $1$ & ${\bf 1_0}^*$
 \\
\hline \hline $P+5V_+$ & $~\chi~$ & $(N^L)_j$ & ${\cal P}_5(f_+)$
& SM
\\
\hline
 $\left(\frac{1}{12},\frac{1}{12},\frac{1}{12};
\underline{\frac{-7}{12},\frac{5}{12}};
\frac{-1}{12},\frac{7}{12},\frac{3}{12}\right)
\left(\frac{3}{12},\frac{1}{12},\frac{-8}{12};0^5\right)'$ & R & $0$
& $1$ & $\overline{\delta}^{-1/6~*}$
\\
$\left(\frac{1}{12},\frac{1}{12},\frac{1}{12};
\frac{-7}{12},\frac{-7}{12};
\frac{-1}{12},\frac{-5}{12},\frac{3}{12}\right)
\left(\frac{3}{12},\frac{1}{12},\frac{-8}{12};0^5\right)'$ & R & $0$
& $1$ & $\overline{\xi}^{-2/3~*}$
\\
$\left(\frac{-5}{12},\frac{-5}{12},\frac{-5}{12};
\frac{-1}{12},\frac{-1}{12};
\frac{5}{12},\frac{1}{12},\frac{-3}{12}\right)
\left(\frac{-9}{12},\frac{1}{12},\frac{4}{12};0^5\right)'$ & R & $0$
& $1$ & $\overline{\eta}^{1/3~*}$
\\
$\left(\frac{-5}{12},\frac{-5}{12},\frac{-5}{12};
\frac{-1}{12},\frac{-1}{12};
\frac{5}{12},\frac{1}{12},\frac{-3}{12}\right)
\left(\frac{3}{12},\frac{1}{12},\frac{-8}{12};0^5\right)'$ & R &
$1_1$ & $1$ & $\overline{\eta}^{1/3~*}$
\\
$\left(\frac{1}{12},\frac{1}{12},\frac{1}{12};
\frac{5}{12},\frac{5}{12};
\frac{-1}{12},\frac{-5}{12},\frac{3}{12}\right)
\left(\frac{-9}{12},\frac{1}{12},\frac{4}{12};0^5\right)'$ & R &
$1_1$ & $1$ & $\overline{\eta}^{1/3~*}$
\\
$\left(\frac{1}{12},\frac{1}{12},\frac{1}{12};
\frac{5}{12},\frac{5}{12};
\frac{-1}{12},\frac{-5}{12},\frac{3}{12}\right)
\left(\frac{3}{12},\frac{1}{12},\frac{-8}{12};0^5\right)'$ & R &
$2_1$ & $1$ & $\overline{\eta}^{1/3~*}$
\\
\hline \hline $P+5V_-$ & $~\chi~$ & $(N^L)_j$ & ${\cal P}_5(f_-)$
& SM
\\
\hline $\left(\frac{-1}{12},\frac{-1}{12},\frac{-1}{12};
\underline{\frac{7}{12},\frac{-5}{12}};
\frac{9}{12},\frac{1}{12},\frac{1}{12}\right)
\left(\frac{3}{12},\frac{5}{12},\frac{-4}{12};0^5\right)'$ & R & $0$
& $1$ & $\delta^{1/6~*}$
\\
$\left(\frac{-1}{12},\frac{-1}{12},\frac{-1}{12};
\frac{7}{12},\frac{7}{12};
\frac{-3}{12},\frac{1}{12},\frac{1}{12}\right)
\left(\frac{3}{12},\frac{5}{12},\frac{-4}{12};0^5\right)'$ & R &
$2_1$ & $1$ & $\xi^{2/3~*}$
\\
$\left(\frac{-1}{12},\frac{-1}{12},\frac{-1}{12};
\frac{-5}{12},\frac{-5}{12};
\frac{-3}{12},\frac{1}{12},\frac{1}{12}\right)
\left(\frac{-9}{12},\frac{-7}{12},\frac{-4}{12};0^5\right)'$ & R &
$0$ & $1$ & $\eta^{-1/3~*}$
\\
$\left(\frac{-1}{12},\frac{-1}{12},\frac{-1}{12};
\frac{-5}{12},\frac{-5}{12};
\frac{-3}{12},\frac{1}{12},\frac{1}{12}\right)
\left(\frac{3}{12},\frac{-7}{12},\frac{8}{12};0^5\right)'$ & R &
$1_1$ & $1$ & $\eta^{-1/3~*}$
\\
$\left(\frac{5}{12},\frac{5}{12},\frac{5}{12};
\frac{1}{12},\frac{1}{12};
\frac{3}{12},\frac{-5}{12},\frac{-5}{12}\right)
\left(\frac{3}{12},\frac{5}{12},\frac{-4}{12};0^5\right)'$ & R &
$1_1$ & $1$ & $\eta^{-1/3~*}$
\\
$\left(\frac{-1}{12},\frac{-1}{12},\frac{-1}{12};
\frac{-5}{12},\frac{-5}{12};
\frac{-3}{12},\frac{1}{12},\frac{1}{12}\right)
\left(\frac{3}{12},\frac{5}{12},\frac{-4}{12};0^5\right)'$ & R &
$1_{\bar{2}}, 4_1$ & $1+1$ & $2\cdot \eta^{-1/3~*}$
\\
\hline
\end{tabular}
\end{center}
\caption{Chiral matter fields satisfying $\Theta=0$ in the $T_5^0$
and $T_5^\pm$ sectors. They are all the right-handed states. Their
${\cal CTP}$ conjugates with the left-handed chirality are provided
from the $T_7^0$, $T_7^+$, and $T_7^-$ sectors. }\label{tb:T5}
\end{table}


\section{Anomalies}\label{App:B}

The anomalies associated with the non-Abelian gauge groups turn
out to be
\begin{eqnarray}
&&  {\rm Tr}[({\rm Non Abel.})^2\cdot {Y}] = {\rm Tr}[({\rm Non
Abel.})^2\cdot Q_{6}]=0
\\
&&  {\rm Tr}[({\rm Non Abel.})^2\cdot Q_{1}] = {\rm Tr}[({\rm Non
Abel.})^2\cdot Q_{2}]
 = {\rm Tr}[({\rm Non Abel.})^2\cdot Q_{3}]
\nonumber \\
&& ~~~~~~~~~~~~~~~~~~~~~~~~~~~ = {\rm Tr}[({\rm Non Abel.})^2\cdot
Q_{4}]=\textstyle-\frac12
\\
&&   {\rm Tr}[({\rm Non Abel.})^2\cdot Q_{B-L}] = {\rm Tr}[({\rm
Non Abel.})^2\cdot Q_{5}]=\textstyle+\frac12 ,
\end{eqnarray}
where ${\rm Non Abel.}=$ ${\rm SU(3)}_c$, ${\rm SU(2)}_L$, and
${\rm SO(10)'}$. U(1)$^3$ type anomalies are
\begin{eqnarray}
&& {\rm Tr}[(Q_Y)^3]={\rm Tr}[(Q_Y)^2\cdot Q_6]=0
\\
&& {\rm Tr}[(6 Q_Y)^2\cdot Q_1]={\rm Tr}[(6 Q_Y)^2\cdot Q_2]={\rm
Tr}[(6 Q_Y)^2\cdot Q_3] \nonumber \\
&& ~~~~~~~~~~~~~~~~~~~~ = {\rm Tr}[(6 Q_Y)^2\cdot Q_4]=-30
\\
&&  {\rm Tr}[(6 Q_Y)^2\cdot  Q_{B-L}]={\rm Tr}[(6 Q_Y)^2\cdot
Q_5]=+30 ,
\end{eqnarray}
and
\begin{eqnarray}
&& {\rm Tr}[(Q_6)^3]={\rm Tr}[(Q_6)^2\cdot Q_Y]=0
\\
&& {\rm Tr}[(Q_6)^2\cdot Q_1]={\rm Tr}[(Q_6)^2\cdot Q_2]={\rm
Tr}[(Q_6)^2\cdot Q_3] \nonumber \\
&& ~~~~~~~~~~~~~~~~~~ = {\rm Tr}[(Q_6)^2\cdot Q_4]=-4
\\
&&  {\rm Tr}[(Q_6)^2\cdot Q_{B-L}]={\rm Tr}[(Q_6)^2\cdot Q_5]=+4 ,
\end{eqnarray}
and so on.

Thus, the anomaly free U(1) charge operators are $Q_Y$, $Q_6$, and
\begin{eqnarray}
&& Q_a = Q_1-Q_2 ,
\\
&& Q_b = Q_1 + Q_2 - 2Q_3 ,
\\
&& Q_c = Q_1 + Q_2 + Q_3 - 3Q_4 ,
\\
&& Q_d = Q_1 + Q_2 + Q_3 + Q_4 + 4Q_5 ,
\\
&& Q_e =\textstyle Q_1 + Q_2 + Q_3 + Q_4 - Q_5 -\frac16 X .
\end{eqnarray}
 The anomalous ${U(1)_{A}}$ is given by
\begin{eqnarray}
Q_A = Q_1 + Q_2 + Q_3 + Q_4 - Q_5 +6X .
\end{eqnarray}
It can be shown that the gravitational anomalies are ${\rm
Tr}Q_Y={\rm Tr}Q_6={\rm Tr}Q_a={\rm Tr}Q_b={\rm Tr}Q_c={\rm
Tr}Q_d={\rm Tr}Q_e=0$, and ${\rm Tr}Q_A= -50$. It can be cancelled
via the Green-Schwarz mechanism~\cite{Green:1984sg}.


\end{document}